\documentclass[submission, Phys]{SciPost}

\hypersetup{
    colorlinks,
    linkcolor={red!50!black},
    citecolor={blue!50!black},
    urlcolor={blue!80!black}
}

\usepackage[bitstream-charter]{mathdesign}
\usepackage{multirow}
\usepackage[normalem]{ulem}

\DeclareSymbolFont{usualmathcal}{OMS}{cmsy}{m}{n}
\DeclareSymbolFontAlphabet{\mathcal}{usualmathcal}

\usepackage{subcaption}

\begin{document}

\begin{center}{\Large 
Exploring Optimal Transport for Event-Level Anomaly Detection at the Large Hadron Collider
}\end{center}

\begin{center}
Nathaniel Craig\textsuperscript{1,2},
Jessica N. Howard\textsuperscript{1,2$\star$} and
Hancheng Li\textsuperscript{1}
\end{center}

\begin{center}
{\bf 1} Department of Physics, University of California, Santa Barbara, CA USA
\\
{\bf 2} Kavli Institute for Theoretical Physics, Santa Barbara, CA USA
\\
${}^\star$ {\small \sf jnhoward@kitp.ucsb.edu}
\end{center}

\begin{center}
\today
\end{center}


\section*{Abstract}
{\bf

Anomaly detection is a promising, model-agnostic strategy to find physics beyond the Standard Model.
State-of-the-art machine learning methods offer impressive performance on anomaly detection tasks, but interpretability, resource, and memory concerns motivate considering a wide range of alternatives.
We explore using the 2-Wasserstein distance from optimal transport theory, both as an anomaly score and as the input to interpretable machine learning methods,
for event-level anomaly detection at the Large Hadron Collider. 
The choice of ground space plays a key role in optimizing performance.
We comment on the feasibility of implementing these methods in the L1 trigger system. 
}

\vspace{10pt}
\noindent\rule{\textwidth}{1pt}
\tableofcontents\thispagestyle{fancy}
\noindent\rule{\textwidth}{1pt}
\vspace{10pt}

\section{Introduction}
\label{sec:intro}
In the search for new physics Beyond the Standard Model (BSM), it is becoming increasingly important to consider model-agnostic search strategies in addition to standard, dedicated searches for a particular model of new physics.
The goal of such broad strategies is to reliably recognize any anomalous data signature which deviates from the Standard Model (SM) baseline. 
These anomaly detection strategies could then be applied to real data, either at the trigger level or in offline analyses, to look for many BSM models at once. 
This is appealing given the plethora of possible BSM signatures and the immense resource requirements of dedicated searches.

Anomaly detection, fueled by state-of-the-art Machine Learning (ML) methods and several Large Hadron Collider (LHC) data challenges~\cite{Govorkova:2021hqu, Aarrestad:2021oeb, Kasieczka:2021xcg}, has increasingly grown in popularity since the discovery of the Higgs boson~\cite{higgs_ATLAS, higgs_CMS} over a decade ago.  
These methods are even starting to be adopted in experimental searches and trigger algorithms~\cite{ATLAS:2023azi, ATLAS:2020iwa, Govorkova:2021utb, Zipper:2023ybp, asres2023spatiotemporal, Grosso:2023owo, CMSECAL:2023fvz, Harilal:2023smf, Chekanov:2023uot}. 
For in-depth reviews see Refs.~\cite{Nachman:2020ccu, hepmllivingreview, Belis:2023mqs}.
When designing anomaly detection algorithms, the types of LHC data considered can broadly be split into three categories. The first considers analyzing the 4-momenta of jet constituents individually or as calorimeter deposits,
while the second considers high-level, reconstructed quantities such as invariant masses and jet substructure variables.
The third operates at the event level considering the 4-momenta of reconstructed objects
and may either correspond to offline or online, trigger-level analyses.
Numerous ML methods have been specially developed,
or adapted from classification tasks, 
for anomaly detection.
Typically, the majority of anomaly detection methods are unsupervised,
so that they may be trained on real, control data, thereby side-stepping biases from simulations. 
However, many methods instead fall under the umbrella of less-than-supervised~\cite{Nachman:2020ccu} (i.e. semi-supervised or weakly supervised).
Additionally, some works~\cite{Dillon:2023zac, Schuhmacher:2023pro} have employed supervised learning in an unsupervised setting. 
Instead of using simulated signal data, they instead transform background data (either real or simulated) so that it appears signal-like.  This ``anomaly augmented" background data is then used as the signal data for supervised training.
In general, ML-based methods offer impressive performance on anomaly detection tasks, far surpassing standard anomaly detection methods, such as kinematic or particle multiplicity cuts~\cite{Nachman:2020ccu, Belis:2023mqs}.

However, this improved performance comes with some drawbacks.
For example, autoencoders, one of the most popular unsupervised anomaly detection methods first used in high-energy contexts in Refs.~\cite{Heimel:2018mkt, Farina:2018fyg}, have been shown to have topological obstructions~\cite{Batson:2021agz}, are prone to high false-negative rates~\cite{gong2019memorizing}, and have difficulty identifying low-complexity anomalies~\cite{Finke:2021sdf,Dillon:2021nxw}. 
Much work has gone into addressing such concerns and developing unbiased, ML-based anomaly detection algorithms~\cite{Dillon:2022mkq, Dillon:2021nxw, Nachman:2020ccu, Freytsis:2023cjr, Fanelli:2023lmp, Belis:2023mqs, Finke:2023ltw, Golling:2023yjq, ATLAS:2023ixc, ATLAS:2023azi, Mikuni:2023tok, Fanelli:2022xwl}. 
However, given the opaque nature of ML methods, the risk of additional, unknown failure modes may still remain.
Moreover, despite recent progress with autoencoder methods~\cite{Govorkova:2021utb, Zipper:2023ybp}, the ease of deploying state-of-the art ML methods, particularly in an online setting where storage space and computing resources are limited, remains a concern~\cite{Govorkova:2021hqu}. 
Furthermore, for many ML methods, it can be difficult to interpret learned relationships and quantify uncertainties~\cite{Belis:2023mqs}.%
\footnote{We note that recent work~\cite{Fanelli:2023lmp} has made great strides towards addressing uncertainty quantification for a particular ML model. Their model ELUQuant uses normalizing flows to estimate the posteriors of a physics-informed Bayesian Neural Network and is capable of capturing both aleatoric and epistemic uncertainties.}
It is therefore useful to explore alternative strategies such as new representations of data that facilitate anomaly detection, either in conjunction with interpretable ML algorithms or without ML altogether.

The theory of Optimal Transport (OT) offers a variety of such promising representations.
OT theory is a branch of mathematics which defines a distance metric between probability distributions. 
Given that the core idea of anomaly detection is to differentiate between SM and BSM distributions over detector signatures, using OT distances as an anomaly score (or as input to distance-based ML algorithms) is sensible. Unlike density estimation strategies~\cite{Nachman:2020ccu}, OT does not need to explicitly learn a probability model of the data. Since OT is predisposed to detect distributional differences, one only needs to define the reference background distribution. Doing so then implicitly defines a probability model of the data.%
\footnote{We note that the particular form of this model will, in general, vary from task to task and thus must be interpreted accordingly to derive precise meaning.}
Therefore, using OT distances as an anomaly score does not require training, is directly interpretable, and is amenable to uncertainty analysis. 
To-date, various OT distances have been widely used in LHC classification tasks~\cite{Larkoski:2023qnv, Cai:2021hnn, Cai:2020vzx, Komiske:2020qhg, Komiske:2019fks, Algren:2023spv, CrispimRomao:2020ejk, Davis:2023lxq, Ba:2023hix, Gaertner:2023tzv}. 
Some previous works have also explored using OT for anomaly detection on jet constituent~\cite{Fraser:2021lxm, Park:2022zov} and high-level variable~\cite{CrispimRomao:2020ejk} data. 
Ref.~\cite{Fraser:2021lxm} investigated using p-Wasserstein distances to augment the latent space of autoencoders, directly as an anomaly score, and in combination with interpretable ML methods.  
Ref.~\cite{Park:2022zov} created a neural embedding of the Energy Mover's Distance (a modified version of the 1-Wasserstein distance) between jets.
Ref.~\cite{CrispimRomao:2020ejk} modified the ground space of the Energy Mover's Distance and used it as an anomaly score on high-level variables.

In this work, we investigate using balanced OT for anomaly detection on L1 trigger-level data from the CMS anomaly detection challenge dataset (CMS ADC 2021)~\cite{Govorkova:2021hqu}.  
Because of its advantageous mathematical properties~\cite{Cai:2021hnn}, we focus on the 2-Wasserstein distance. 
We use it both directly as an anomaly score as well as in tandem with two interpretable ML techniques: one-class Support Vector Machines (SVMs)~\cite{scholkopf1999support} and k-Nearest Neighbors (kNNs)~\cite{1053964} trained using anomaly augmented background as signal.
Thus, the methods we consider range from unsupervised to weakly supervised.
We find that, unsurprisingly~\cite{Kasieczka:2022naq}, the choice of ground space plays a large role in performance.
By considering a different ground space from previous works~\cite{Gaertner:2023tzv, Ba:2023hix, Davis:2023lxq, Cai:2021hnn, Cai:2020vzx, Komiske:2020qhg, Komiske:2019fks}, we are able to significantly improve performance. We compare our results to baseline methods and a recent state-of-the-art ML method~\cite{Dillon:2023zac} benchmarked on the same dataset. We conclude with comments about the feasibility of incorporating these methods into the Level-1 (L1) trigger system and directions for future work to improve performance.

\section{Dataset}
\label{sec:dataset}

In this work we use simulated event-level anomaly data from the CMS anomaly detection challenge dataset (CMS ADC 2021)~\cite{Govorkova:2021hqu}, which considers proton-proton collisions at a center-of-mass-energy of 13~{\rm TeV} simulated using {\sc Pythia} 8.240~\cite{Sjostrand:2014zea}, and {\sc Delphes} 3.3.2~\cite{deFavereau:2013fsa} with the Phase-II CMS detector card. 
There are six sub-datasets: a background dataset comprising only SM processes, four signal datasets with potential BSM processes for new particles, and a ``mystery'' dataset containing both background and signal events for an unknown signal process.  In this work, we will not consider the latter.

This dataset is restricted so that any event can have at most 4 electrons/photons,\footnote{The dataset does not distinguish between electrons and photons.} 4 muons, and 10 jets. There is also one aggregate missing transverse energy (MET) object per event. For each event, the $p_{\rm T}/{\rm GeV}, \eta, \phi$, and identity of each object is stored in a (19,4), zero-padded array.
Events are also required to have at least one electron/photon or muon with $p_{\rm T}>23~{\rm GeV}$. Additionally, all electrons/photons must have $p_{\rm T}>3~{\rm GeV}$ and $|\eta|<3$; all muons must have $p_{\rm T}>3~{\rm GeV}$ and $|\eta|<2.1$; all jets must have $p_{\rm T}>15~{\rm GeV}$ and $|\eta|<4$ and are reconstructed using {\sc FastJet}~\cite{Cacciari:2011ma} with the anti-$k_t$ algorithm~\cite{Cacciari:2008gp}.
This data format is meant to mimic the kind of information available at the LHC's L1 trigger level, where bandwidth, latency, and resources are limited.

The SM background data contains four leading processes with representative fractions determined by their leading-order cross section and trigger efficiency: 
\begin{align}
    &pp \rightarrow W^\pm + {\rm jets} \rightarrow l^\pm \nu_l + {\rm jets} &(59.2\%)~ \nonumber\\ 
    &pp \rightarrow {\rm jets} &(33.8\%)~\nonumber\\
    &pp \rightarrow Z + {\rm jets} \rightarrow l^+ l^- + {\rm jets} &(6.7\%)~ \nonumber\\
    &pp \rightarrow t\bar{t} + {\rm jets} &~(0.3\%),
\end{align}
where $l = e, \mu, \tau$. Four BSM signal models are also considered:
\begin{itemize}
    \item \textbf{Neutral scalar boson $A$}: The mass of this BSM particle is $m_A = 50 ~{\rm GeV}$. The production mechanism is $pp \rightarrow A + X \rightarrow Z^* Z^* + X, ~~Z^* \rightarrow l^+ l^-$, where $X$ indicates inclusive activity. 
    \item \textbf{Scalar boson $h^0$}: The mass of this BSM particle is $m_{h^0} = 60 ~{\rm GeV}$. The production mechanism is $pp \rightarrow h^0 + X \rightarrow \tau^+ \tau^- + X$, where $X$ indicates inclusive activity.
    \item \textbf{Charged scalar $h^\pm$}: The mass of this BSM particle is $m_{h^\pm} = 60 ~{\rm GeV}$. The production mechanism is $pp \rightarrow h^\pm + X \rightarrow \tau^\pm v + X$, where $X$ indicates inclusive activity.
    \item \textbf{Leptoquark ($LQ$)}: The mass of this BSM particle is $m_{LQ} = 80 ~{\rm GeV}$. The production mechanism is $pp \rightarrow LQ \rightarrow \tau b$.
\end{itemize}

As was pointed out in Ref.~\cite{Dillon:2023zac, Mikuni:2021nwn}, many standard methods are sensitive to these BSM signals.  Therefore, in order to establish baseline performance, we consider  standard cuts on kinematic quantities: total $p_{\rm T}$, ${\rm MET}$, and total multiplicities of $e/\gamma$, $\mu$, ${\rm jet}$, respectively.  The performance of these baseline quantities are reported in Table~\ref{tab:baseline_performance} of Appendix~\ref{app:baseline_performance}.

\subsection{Anomaly augmentations}
\label{sec:anomaly_aug}
In addition to fully unsupervised methods, we also consider using supervised learning in an unsupervised setting.  
To accomplish this, we apply the anomaly augmentations described in Ref.~\cite{Dillon:2023zac} to the SM background data described in the previous section. 
This creates a "fake" signal dataset which we can subsequently use in supervised training.

We consider three types of anomaly augmentations~\cite{Dillon:2023zac}:
\begin{enumerate}
    \item \textbf{Multiplicity increase:} This transformation randomly adds some number of $e/\gamma$, $\mu$, and ${\rm jets}$ to each event in the range $(n_{e/\gamma}, 4-n_{e/\gamma})$, $(n_{\mu}, 4-n_{\mu})$, $(n_{\rm jets}, 10-n_{\rm jets})$, respectively. The $p_T$ of each new particle is a randomly chosen fraction of the highest $p_T$ of any object in that event added to the base required by that object's selection criteria. The $\eta,\phi$ of each object are uniformly chosen within the allowed limits corresponding to that object's type. After all objects are added, the ${\rm MET}$ of the event is recalculated.
    \item \textbf{Multiplicity increase, with constant ${\rm MET}$ and $p_{\rm T}$:} This transformation keeps the total $p_{\rm T}$ and ${\rm MET}$ constant while still increasing the overall object multiplicity.  This is achieved by splitting an existing object to create two new objects. The combined $p_{\rm T}$ of the two new objects is equal to the original.  The $\eta,\phi$ of each new object are then randomly smeared with Gaussian noise.
    \item \textbf{${\rm MET}$ and $p_{\rm T}$ shift:} This transformation randomly shifts the a) ${\rm MET}$, b) reconstructed object $p_{\rm T}$, or c) both by a constant multiplicative factor. Transformations (a),(b),(c) are chosen with equal probability.  To satisfy selection criteria, the multiplicative factor for shifting $p_{\rm T}$ is chosen uniformly in the range $[1,5)$ (i.e. the $p_{\rm T}$ of objects will never be down-shifted such that it might violate an object's minimum $p_{\rm T}$ criteria).  No such restriction exists for ${\rm MET}$, so its multiplicative factor is chosen uniformly in the range $[0.5,5)$.
\end{enumerate}
In general, we apply augmentations (1), (2), and (3) with equal probability. However, there are certain events for which augmentation (2) could not transform the event without causing the event selection criteria to be violated (approximately 8\%). 
For these events, we instead apply augmentations (1) and (3) with equal probability. Taking this into account, approximately 34.6\%, 30.6\%, 34.6\% of events are transformed with augmentations (1), (2), and (3), respectively.

\section{Review of Optimal Transport (OT) distances}
\label{sec:OT_dist_review}

OT theory provides a framework for comparing probability distributions by assigning a cost to transform one probability distribution into another. 
    The cost of moving a differential piece of the source distribution to construct the target distribution is the amount of probability mass weighted by (an integer power of) how far that differential piece has to move in the ground space. 
    The collection of these differential movements is called a transport plan, and the net cost of executing the {\it optimal} (lowest-cost) transport plan induces a distance between the source and target distributions. 
    In this way, OT theory gives a way to elevate cost metrics defined on the ground space to a metric between probability distributions defined over that ground space.
  
    A mathematically robust choice of ground space cost metric is the squared Euclidean distance, $c(x, x') = |x - x'|^2$.  
    The distance associated with the solution of the OT problem with this cost metric is the 2-Wasserstein distance.   
    Let $\mathcal{E}$ ($\mathcal{E}'$) be a discrete source (target) distribution over a $d$-dimensional ground space, $X$. Therefore, $\mathcal{E}$ ($\mathcal{E}'$) is a collection of probability point masses, $p_i$ ($p'_j$), at locations $x_i\in X$ ($x'_j \in X$).
    The 2-Wasserstein distance is then given by
    \begin{align}
        W_2(\mathcal{E},\mathcal{E}') = \left( \underset{g \in \Gamma(\mathcal{E},\mathcal{E}')}{\rm min}  \sum_{i,j} g_{ij} |x_i - x'_j|^2 \right)^{1/2}
    \end{align}
    where $\Gamma(\mathcal{E},\mathcal{E}')$ is the set of possible joint distributions with marginals $\mathcal{E},\mathcal{E}'$, respectively. The $g^* \in \Gamma(\mathcal{E},\mathcal{E}')$ which minimizes the above is the optimal transport plan. 
       
    While many works have employed variations on the 1-Wasserstein distance (or {\it Earth Mover's Distance}) due to its conceptual simplicity~\cite{Gaertner:2023tzv, Larkoski:2023qnv, Ba:2023hix, Davis:2023lxq, Komiske:2020qhg, CrispimRomao:2020ejk, Komiske:2019fks}, using the alternative 2-Wasserstein structure is mathematically beneficial for several reasons.  
    The strictly convex cost function guarantees that $g^*$ is unique, admits unique geodesics between distributions, and lends itself to efficient linearization algorithms leveraging the resulting pseudo-Riemannian structure~\cite{Cai:2020vzx}. 
    Therefore, in this work, we will only consider the 2-Wasserstein distance.
    Having fixed this choice, we are left to define the underlying ground space, $X$, and the distributions over it. In this work, we consider several possible choices (see Section~\ref{sec:ground_space_choice}).

\section{Choice of ground space distance}
\label{sec:ground_space_choice}

\begin{figure}
    \centering
    \includegraphics[width=\textwidth]{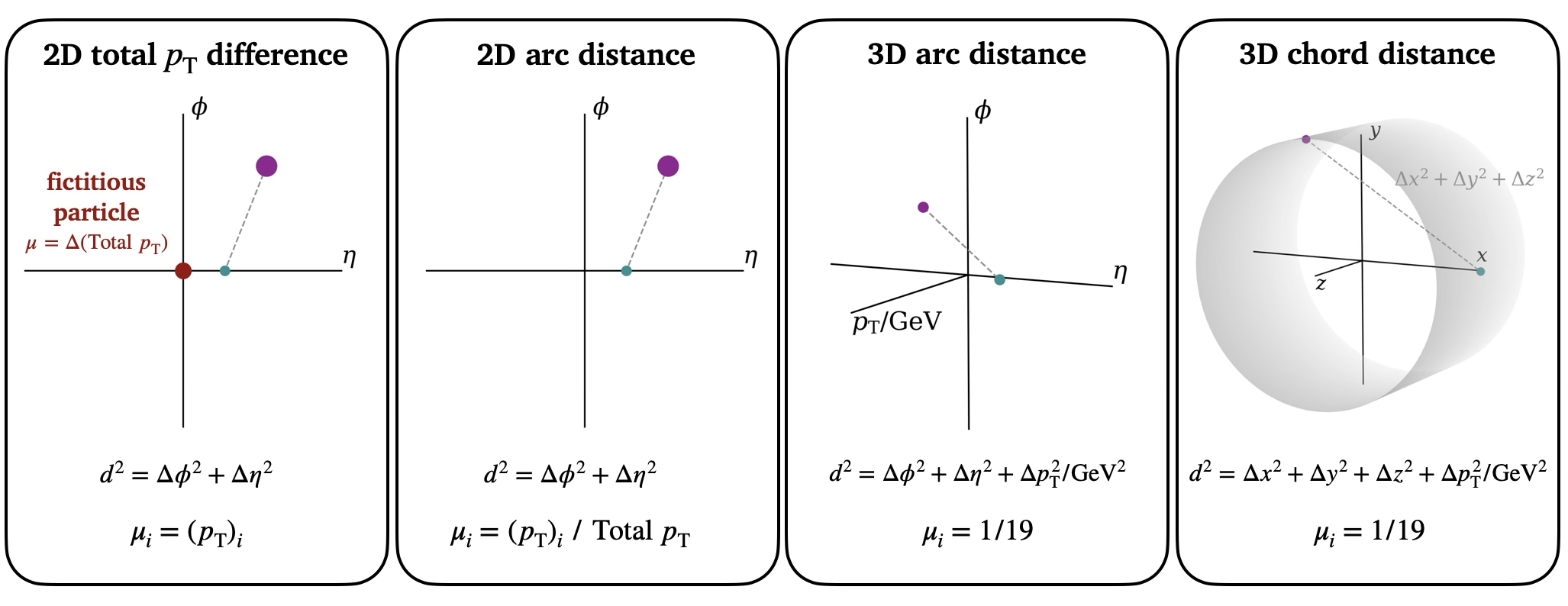}

    \caption{Cartoons describing different ground space choices. Each panel shows the difference between two points in the ground space ($d^2$) as well as the probability weighting of the $i^{th}$ point in an event ($\mu_i$). 
    }
    \label{fig:ground_space_choice}
\end{figure}

Given that probability distributions are sensitive to the choice of ground space and coordinate transformations~\cite{Kasieczka:2022naq}, 
the 2-Wasserstein distance is sensitive as well.  
This means that selecting a ground space that maximizes the differences between SM and BSM events (and minimizes differences between SM events) is essential. 
In this section, we describe the various choices of ground space distances considered in this work.
These options are summarized in Figure~\ref{fig:ground_space_choice}.

\subsection{2D distances}
Past works~\cite{Fraser:2021lxm, Gaertner:2023tzv, Ba:2023hix, Davis:2023lxq, Cai:2021hnn, Cai:2020vzx, Komiske:2020qhg, Komiske:2019fks} have opted for a physically-motivated ground space where the relative $\eta, \phi$ plane%
\footnote{Technically, this is actually a cylinder due to the periodic nature of $\phi$. However, the cylinder is often flattened and referred to as a plane in the literature.}
is centered on a jet's 4-vector location. 
The (discrete) probability distribution is proportional to the $p_{\rm T}$ associated to each jet constituent in this plane. 
However, this intuitive picture has drawbacks when considering scaling OT methods from a jet-level to an event-level picture. 

For balanced OT, the distributions must be identically normalized and thus information about the total $p_{\rm T}$ associated with an event is lost. While this information may be regained by moving to an unbalanced OT framework~\cite{CrispimRomao:2020ejk, Cai:2021hnn}, the many possible unbalanced OT prescriptions come with varying degrees of mathematical well-posedness and typically entail a significant increase in computational cost~\cite{Cai:2021hnn}.
Additionally, the underlying physical symmetries of collider events motivate  aligning distributions in the $\eta, \phi$ plane of the detector before computing OT distances. Unlike with jets, accounting for such symmetries by pre-processing the data is typically infeasible for full events, causing physically identical distributions to appear different. 

In this work, we are interested in constructing alternative ground spaces to circumvent many of these difficulties.
We consider two ``2D'' baseline strategies where the ground space is chosen to be the event-level $\eta, \phi$ plane. 
The first strategy, \textbf{2D arc distance}, is the classic balanced OT strategy. Each event is a collection of 19 points in the 2D $\eta,\phi$ plane. They are each given a probability weighting $\mu_i = (p_{\rm T})_i ~/~ {\rm Total} ~p_{\rm T}$. The ground space distance between two points is then $d^2 = \Delta \phi^2 + \Delta \eta^2$. 

The second strategy, \textbf{2D total $p_{\rm T}$ difference}, uses the same ground space choice, but adds a fictitious particle at the origin to absorb total $p_{\rm T}$ differences between events. Thus, an event now has 20 points -- 19 physical and 1 fictitious. The physical points now have probability weighting $\mu_i = (p_{\rm T})_i$ and the fictitious point has probability weighting $\mu = \Delta ({\rm Total} ~p_{\rm T})$. This restores total momentum information by considering a less physical event representation.

\subsection{3D distances}
Considering alternative ground spaces may offer a way to circumvent or minimize the effect of the difficulties of the 2D distances.  For example,
Ref.~\cite{Larkoski:2023qnv} sought to address many of these shortcomings by considering a spectral ground space that encodes pairwise particle
angles and products of particle energies.

In this work, we consider alternate ``3D'' ground spaces which add an additional $p_{\rm T}/{\rm GeV}$ dimension to the standard $\eta, \phi$ plane. Doing so incorporates  information about the $p_T$ distribution of an event.%
    \footnote{We note that this choice of presentation is analogous to that used by many state-of-the-art machine learning methods.}
To form a distribution over this space, each of the 19 objects is given an equal probability weighting. 
This allows us to use the 2-Wasserstein distance in a balanced OT setting, which maintains its mathematical advantages and admits computationally efficient algorithms.  

Non-existent objects (i.e. zero padded rows in an event) sit at the origin of this space.
However one can choose to define the origin in different ways.%
\footnote{This choice now matters since fictitious particles will no longer have $\mu=0$, so their location matters.}
In the first strategy, \textbf{3D arc distance}, we define the origin to sit on the cylinder's surface at $\eta, \phi = 0,0$. The ground space distance is therefore $d^2 = \Delta p_{\rm T}^2/{\rm GeV^2} + \Delta \phi^2 + \Delta \eta^2$. 
In the second strategy, \textbf{3D chord distance}, we make a somewhat more physical choice and place the origin at the center of the cylinder i.e. $x, ~y, ~z = 0,0,0$ and use chord lengths between points rather than arc lengths i.e. $d^2 = \Delta p_{\rm T}^2/{\rm GeV^2} + \Delta x^2 + \Delta y^2 + \Delta z^2$.%
\footnote{We thank the reviewer for suggesting this approach.}
This way a particle at $\eta, \phi_1$ will be equally far away from the origin as a particle at $\eta, \phi_2$.
The radius of the cylinder, $r$, can be arbitrary, so we choose it so that the 3D arc and 3D chord distances between a real and fictitious particle have approximately the same range.
In particular, we want $d^2_{\max}$ to be approximately the same in both cases. 
This means we want $r$ such that $\max\{\Delta \phi^2 + \Delta \eta^2\} \approx \max\{\Delta x^2 + \Delta y^2 + \Delta z^2\}$. This implies $r^2 = \frac{\pi^2 + 4^2}{1 + {\rm sinh}^2 4} \approx 0.0347$.
This also implies that $d^2_{\min}$ is very similar in both cases as well. 

\section{Methods}
\label{sec:methods}

\begin{figure}[t]
    \centering
    \includegraphics[width=\linewidth]{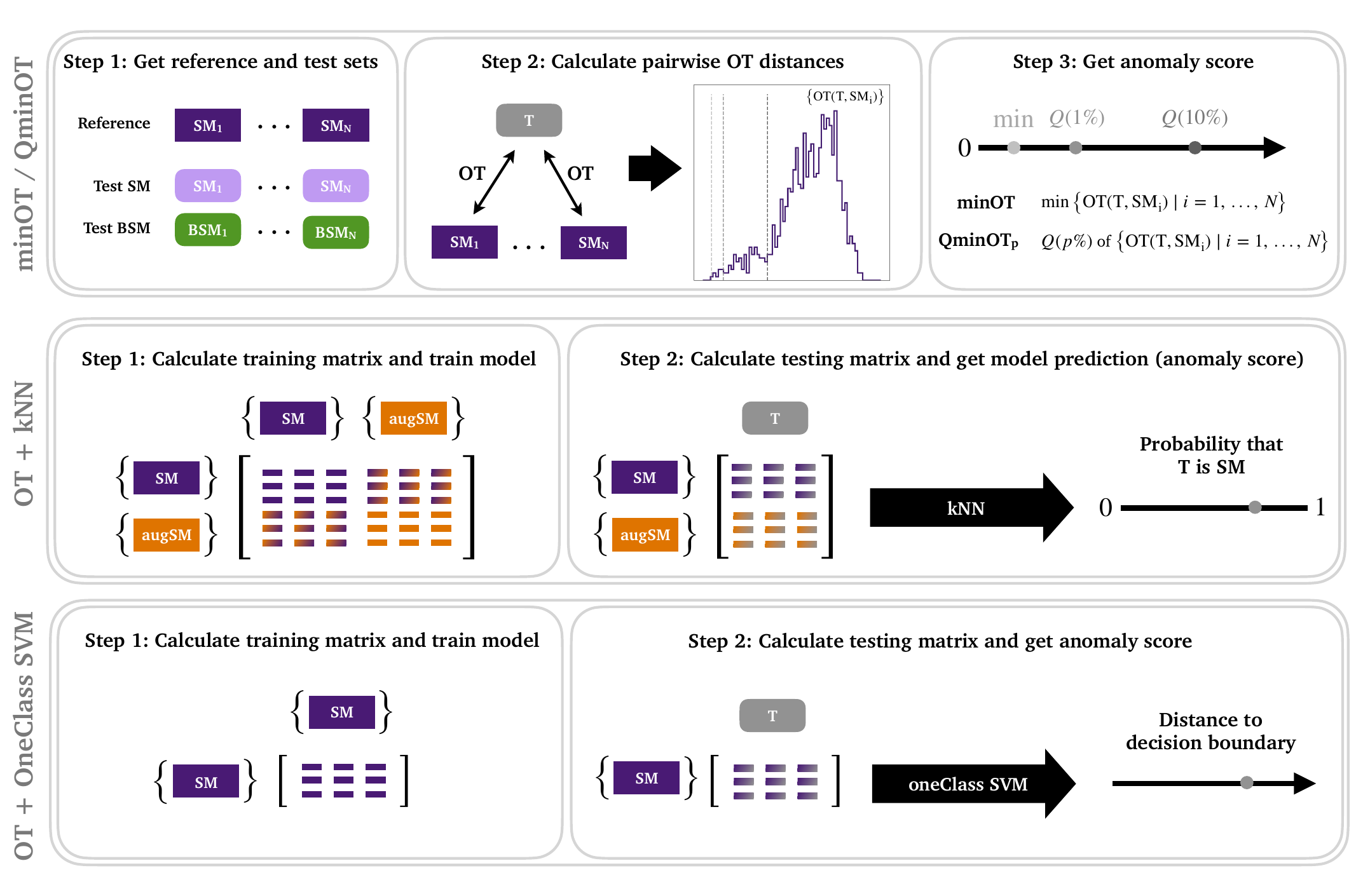}
    \caption{Flowchart of the two classes of methods considered in this work. The first class of methods use the closest OT distance to a reference set of standard model events. Where ``closest'' is defined as the minimum distance (minOT) or the quantile minimum distance (QminOT). The second class of methods use pairwise OT distances as input to machine learning classifiers (kNN or SVM). During training the kNN, anomaly augmented SM events are used as the stand-in for the BSM signals. 
    }
    \label{fig:methods_overview}
\end{figure}

In this section, we describe the details of the methods used in this work. 
We utilize the Python Optimal Transport ({\sc POT}) library~\cite{flamary2021pot} to calculate 2-Wasserstein distances and the {\sc scikit-learn} library~\cite{scikit-learn} to implement the k-Nearest Neighbor and one-class Support Vector Machine algorithms.%
\footnote{The code to reproduce all results and figures in this work is publicly available at \href{https://github.com/hancheng-li/anomaly_detection_code}{https://github.com/hancheng-li/anomaly\_detection\_code}.}
The workflow of the strategies considered are summarized in Figure~\ref{fig:methods_overview}.

    \subsection{Closest Optimal Transport (OT) distance}
    \label{sec:min_OT}
    A key expectation is that the 2-Wasserstein distance between two SM events should generally be less than the distance between a SM event and a BSM event. 
    Thus, we can attempt to use the 2-Wasserstein distance between a test event and a representative ensemble of SM events to derive an anomaly score.
    More precisely, let $\mathcal{T}$ be a test event of unknown type and $\{ \mathcal{S}_1, ..., \mathcal{S}_N\}$ be a collection of $N$ SM reference events.  
    This will, in turn, yield $N$ 2-Wasserstein distance values, $\{ W_2(\mathcal{T},\mathcal{S}_1), ..., W_2(\mathcal{T},\mathcal{S}_N)\}$.  
    We then want to condense these $N$ event-to-event distances into a single event-to-ensemble distance.  
    There are many ways one could envision doing this.  
    Ref.~\cite{Fraser:2021lxm} explored taking the average value of $\{ W_2(\mathcal{T},\mathcal{S}_1), ..., W_2(\mathcal{T},\mathcal{S}_N)\}$, as well as choosing the reference events $\{ \mathcal{S}_1, ..., \mathcal{S}_k\}$ to be the $k$-medoids of the SM distribution (i.e. reducing $N \rightarrow k$) and taking the minimum of $\{ W_2(\mathcal{T},\mathcal{S}_1), ..., W_2(\mathcal{T},\mathcal{S}_k)\}$.
    In this work, we consider two other possibilities for our event-to-ensemble distance.
    
    First, in the \textbf{minOT} approach, we define the anomaly score for a test event $\mathcal{T}$ to be the minimum of the ensemble. Namely, the anomaly score of $\mathcal{T}$ is
    \begin{equation}
        \textbf{minOT}: \quad \text{anomaly score} = \min ~\{ W_2(\mathcal{T},\mathcal{S}_1), ..., W_2(\mathcal{T},\mathcal{S}_N)\}\,.
    \end{equation}
    Using the minimum distance between a test event and an ensemble of SM events as an anomaly score is consistent with the expectation that a given SM event should be closer to at least a subset of the SM background events than a corresponding BSM event, even if the background sample is highly inhomogeneous.\footnote{We have verified that using the maximum distance as an anomaly score is ineffective precisely because of the inhomogeneity of the background sample. Similarly, we have verified that the average distance also suffers from the inhomogeneity of the background sample.} Thus, a large minOT anomaly score implies $\mathcal{T}$ is more likely to be a BSM event.
    
    While fast and simple, the \textbf{minOT} definition may be prone to random fluctuations arising from the finite number of reference events, which may make it unstable. A more stable strategy is to use the quantile minimum OT distance (\textbf{QminOT}).\footnote{We are grateful to the reviewer for suggesting this alternative.} Let $Q$ be the quantile function of the reference ensemble $\{ W_2(\mathcal{T},\mathcal{S}_1), ..., W_2(\mathcal{T},\mathcal{S}_N)\}$. For $p \geq 1/N$, we can then define the $p$ quantile minimum OT distance as
    \begin{equation}
        \textbf{QminOT}_p: \quad \text{anomaly score} = Q(p)\,.
    \end{equation}
    Concretely, for $N=1,000$, ${\rm QminOT}_{1\%}$ would (approximately) be the $10^{\rm th}$ smallest OT distance in the ensemble $\{ W_2(\mathcal{T},\mathcal{S}_1), ..., W_2(\mathcal{T},\mathcal{S}_N)\}$. Note that in the case that $p=1/N$, then the two methods are equivalent up to quantile estimation effects. We use the median unbiased quantile function in the {\sc numpy} library~\cite{Hyndman01111996}.

    For the main results of this work, we construct a reference set from $N=2,000$ SM background events and a test set from another set of $N=2,000$ SM background events and $N=2,000$ BSM signal events (for each signal model). We repeat this process for 5 different random seeds and report the average and standard deviation. Therefore, our results capture uncertainties arising from the particular reference set chosen as well.
    Additionally, Appendix~\ref{app:scaling_with_N} explores how the performance of these methods scale with $N$.

    \subsection{k-Nearest Neighbors (kNNs) using OT distances}
    \label{sec:kNN}

    The kNN algorithm~\cite{1053964} provides a simple and interpretable supervised machine learning strategy for binary classification. Given a training set of events from both classes (SM background and BSM signal), a test event's class is determined by a majority vote from its $k$ nearest neighbors, where $k$ is a hyperparameter. 
    The 2-Wasserstein distance is used to calculate the distance between events to determine which neighbors are closest.

    Using the 2-Wasserstein distance in this way helps take advantage of the tendency for events to cluster instead of fully separate~\cite{Cai:2020vzx, Cai:2021hnn, Fraser:2021lxm, Komiske:2019fks}. The SM background events in our study are highly inhomogeneous, in that they include many different physics processes under the same SM background label. Therefore, it is possible that the differences between two SM subprocesses are more extreme than the differences between a SM subprocess and a BSM process. In this scenario, a clean separation would be unlikely. However, we would still expect events from the same process to be more similar to each other than to events from a different process. 
    This would lead to clusters of events forming in this abstract 2-Wasserstein space. 
    This has been empirically observed in prior works operating on the 2D ground space~\cite{Cai:2020vzx, Cai:2021hnn, Fraser:2021lxm, Komiske:2019fks}. 

    Even though kNNs are a supervised classification algorithm, we can promote them to a weakly supervised anomaly detection method by using anomaly augmented SM events (instead of BSM events) as ``signal'' during training. 
    We employ the anomaly augmentations proposed in Ref.~\cite{Dillon:2023zac}, see Section~\ref{sec:anomaly_aug} for details.
    The training dataset consists of a random mixture of $2{,}000$ Standard Model (SM) background events and $2{,}000$ statistically independent anomaly-augmented background events. We reserve $20\%$ of this dataset for validation and scan the hyperparameter $k$ over the range $[3,63]$ in increments of $2$, selecting the value that yields the best validation performance. The model is then retrained on the full training dataset using the optimal value of $k$. Finally, for each signal model, we construct a test set consisting of a random mixture of $2{,}000$ SM background events and $2{,}000$ beyond-the-Standard-Model (BSM) signal events. Appendix~\ref{app:scaling_with_N} also explores how this method scales with $N$.
    
    Additionally, to gauge how much information is contained in the representations of the data that we consider (i.e. 2D and 3D ground space) we also perform fully supervised classification using kNNs. 
    In some sense, this is a best-case performance limit for anomaly detection, and serves to investigate how the choice of ground space affects performance.
    For each BSM signal model, we train on a random mixture of $2,000$ SM background events and $2,000$ BSM signal events, with $20\%$ reserved for validation and $20\%$ reserved for testing. 
    We scan over the hyperparameter $k$ in the range $[3, 63]$ in increments of $2$ and select the value which yields the best performance. 
    For each classification task, the mean and standard deviation of the results are computed with 5 different sets of independent SM background events and BSM signal events (for each signal model).

    \subsection{One-class Support Vector Machines (SVMs) using OT distances}
    \label{sec:SVM}
    
    One-class SVMs~\cite{scholkopf1999support} are another simple and interpretable machine learning strategy. 
    Unlike their vanilla counterpart (SVMs), one-class SVMs are fully unsupervised and operate by drawing a decision boundary around the training SM data. 
    It therefore implicitly learns to estimate the density of SM events and assign a BSM label to test events falling in the low-density, tail regions. 
    Importantly, the decision boundary is typically encouraged to circumscribe the entire SM density. Therefore, a BSM event which lies in a low-density region but is surrounded by SM events will still fall within the decision boundary and thus be labeled as background.
    Therefore, one-class SVMs are commonly used for ``outlier'' detection tasks and are not usually efficient at detecting clusters within a background distribution.%
    \footnote{While outliers are a certain type of anomaly it has been previously noted that, in high-energy physics contexts, it is more likely for anomalies to be ``over-densities''~\cite{Kasieczka:2021xcg}. This observation generally makes outlier detection methods less favorable for anomaly detection at the LHC.}
    Similar to the kNN case, we distribute events in a high-dimensional, abstract 2-Wasserstein space where we expect like-events to cluster.
    There are two hyperparameters: $\nu$, which controls the margin of the decision boundary, and $\gamma$ which is (inversely) related to the variance of the Gaussian kernel, $K(\mathcal{E},\mathcal{E}') = {\rm exp}\left( - \gamma W_2(\mathcal{E}, \mathcal{E}')^2 \right)$. 
    We will see that the parameter $\gamma$ is  crucial in circumventing the aforementioned issue.
    
   The training dataset for the one-class SVM consists of $2{,}000$ SM background events. We fix the hyperparameter $\nu=0.2$ and reserve $20\%$ of the training data for validation. We then scan $\gamma$ over the range $[0.05,0.55]$ in increments of $0.10$, evaluating each value using the held-out SM background events together with statistically independent BSM validation events from all four signal models. We select a single value of $\gamma$ based on the mean validation performance across five independent subsets. The model is then retrained on the full set of $2{,}000$ SM background events using the selected hyperparameters. Finally, for each signal model, we construct a test set consisting of a random mixture of $2{,}000$ statistically independent SM background events and $2{,}000$ BSM signal events.
  
\section{Results}
\label{sec:results}
\begin{figure}[h]
    \centering
    \includegraphics[width=\textwidth]{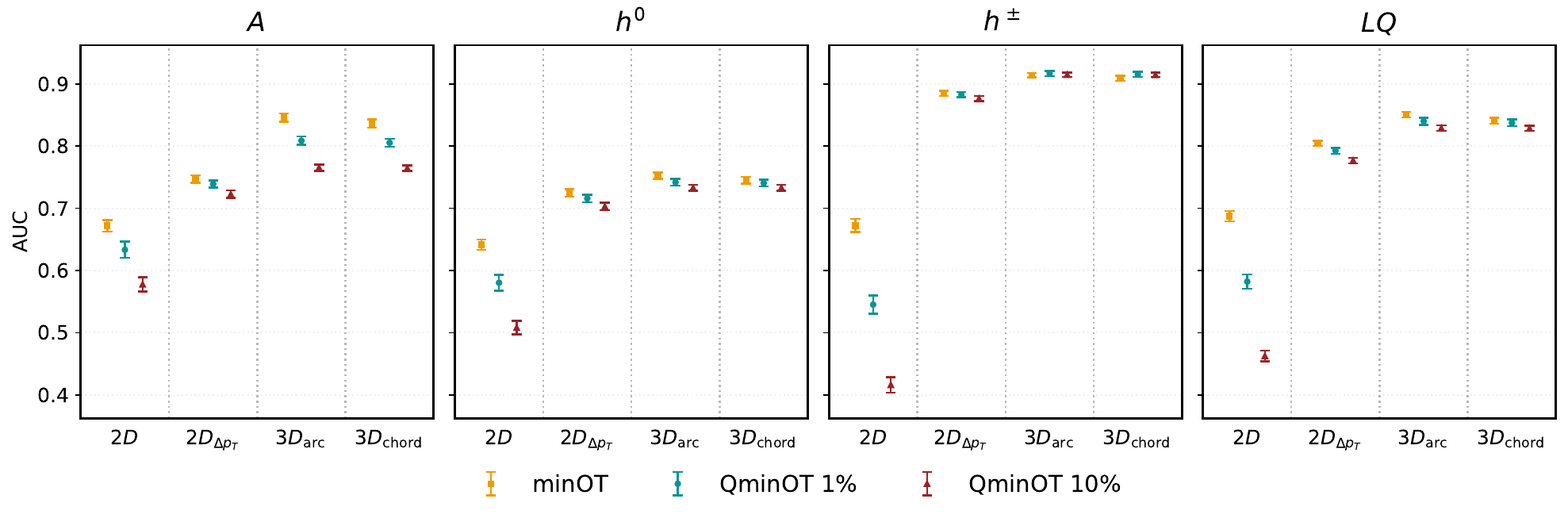}
    \caption{
    AUC comparison of OT-only anomaly scores across choices of ground space distance, shown separately for each signal case.
    Results are given for the minOT, QminOT$_{1\%}$, and QminOT$_{10\%}$ distances on the $2$D, $2$D$_{\Delta p_T}$, $3$D$_{\rm arc}$, and $3$D$_{\rm chord}$ ground spaces.
    Markers denote the mean AUC over five independent repeats with $N=2,000$ test events, and error bars denote the corresponding standard deviation.
    Three features stand out:
    First, the 3D ground spaces substantially outperform the 2D ground space, with $3$D$_{\rm arc}$ and $3$D$_{\rm chord}$ performing almost identically.
    Second, including the $p_T$ difference improves on the plain 2D ground space, but $2$D$_{\Delta p_T}$ still falls short of either 3D ground space.
    Third, QminOT$_p$ and minOT perform comparably at the small quantile fractions considered here.
    }
    \label{fig:OTonly_AUC_comparison}
\end{figure}
We compare performance across different methods with Receiver Operating Characteristic (ROC) curves and several standard metrics derived from ROC curves. 
Namely, we consider the area under the ROC curve (AUC), inverse background efficiency rate ($\epsilon_b^{-1}$), Significance Improvement (SI), and F1 score.
These are summarized for the best performing methods in Table~\ref{tab:OT_anomaly_detection_comparison}.
To improve visualization of differences in the ROC curves of various methods we plot the inverse background efficiency rate ($\epsilon_b^{-1}$) as a function of signal efficiency ($\epsilon_s$).
All anomaly detection methods are evaluated on 5 separate test sets and the resulting ROC curves and metrics are averaged and standard deviation calculated.

\begin{figure}[]
    \centering
    \includegraphics[width=\textwidth]{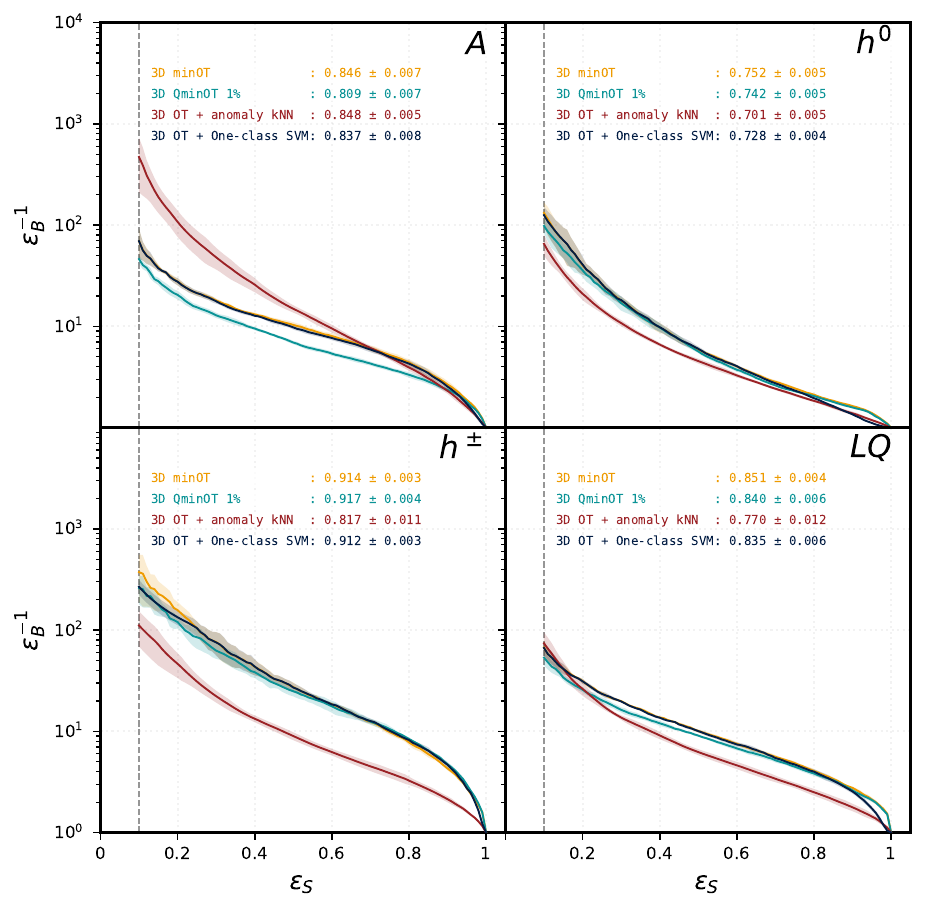}
    \caption{
    ROC curves with corresponding AUC values for each signal case are reported for the anomaly detection methods considered. For the closest optimal transport approach, we show the result for minOT and QminOT$_{1\%}$ distance as an anomaly score as well as using OT distances in tandem with interpretable machine learning methods (3D OT + anomaly kNN, 3D OT + one-class SVM). 
    The 3D minOT, 3D QminOT, and 3D OT + one-class SVM are all unsupervised methods whereas the 3D OT + anomaly kNN is weakly supervised. 
    Note that the vertical axis is the inverse background efficiency rate ($\epsilon_B^{-1}$) instead of background efficiency ($\epsilon_B$). 
    }
    \label{fig:ROC_curves}
\end{figure}

\begin{figure}[]
    \centering
    \includegraphics[width=\textwidth]{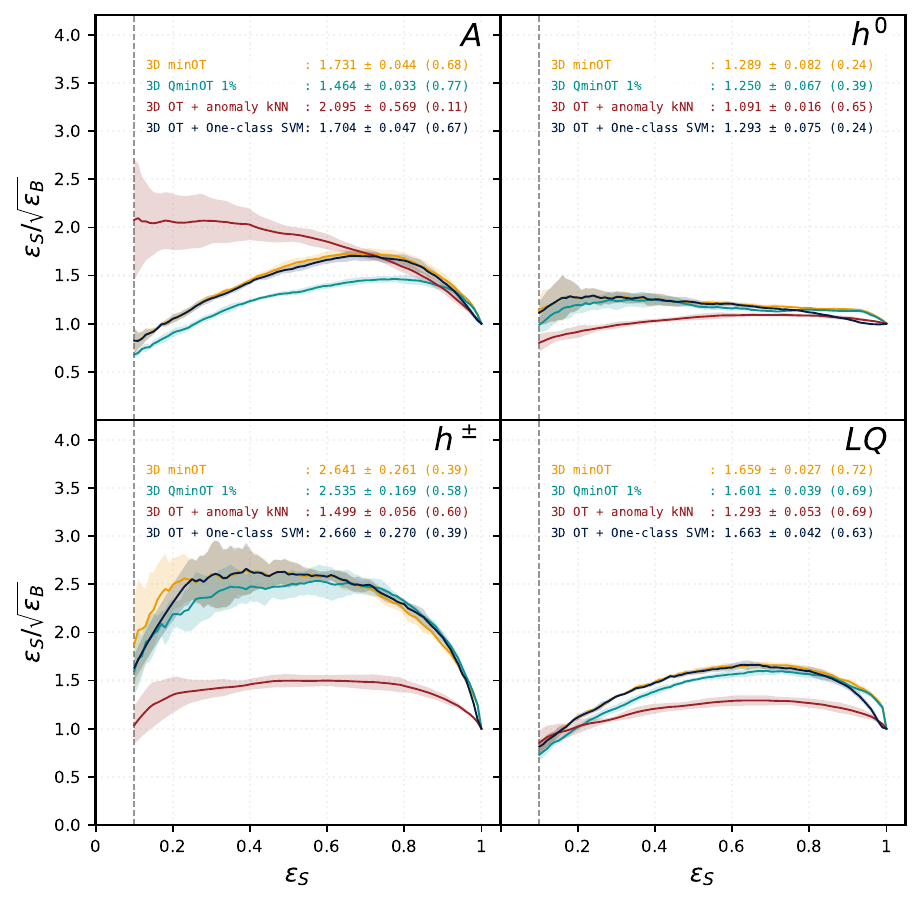}

    \caption{
    SI curves with the corresponding max-SI values are reported for each signal case and each anomaly detection method considered. For the closest optimal transport approach, we show results using the 3D minOT and QminOT$_{1\%}$ distances as anomaly scores, as well as results from using OT distances in tandem with interpretable machine learning methods: 3D OT + anomaly kNN and 3D OT + one-class SVM. The 3D minOT, 3D QminOT$_{1\%}$, and 3D OT + one-class SVM methods are unsupervised, whereas 3D OT + anomaly kNN is weakly supervised. A lower signal-efficiency threshold of $\epsilon_S = 0.1$ is imposed because the SI curves become unstable at low signal efficiencies. The max-SI value is found within this signal-efficiency range, $\epsilon_S \in [0.1, 1]$, and the signal efficiency at which it occurs is reported in parentheses after the max-SI value.}
    \label{fig:SI_curves}
\end{figure}

\begin{table}[ht]
    \centering
    \resizebox{\textwidth}{!}{%
    \begin{tabular}{c c|c|c|c|c|}
    && \multicolumn{4}{c|}{3D OT anomaly-detection performance} \\
    \cline{3-6}
    && \multicolumn{2}{c|}{OT-only methods} & \multicolumn{2}{c|}{OT+ML methods} \\
    \cline{3-6}
    && \rule{0pt}{2ex}3D minOT & 3D QminOT$_{1\%}$ & \begin{tabular}[t]{@{}c@{}}3D OT + anomaly kNN\\[0pt]{$k=17$}\end{tabular} & \begin{tabular}[t]{@{}c@{}}3D OT + one-class SVM\\[1pt]{$\nu=0.2,\ \gamma=0.55$}\end{tabular} \\
    \hline
    \multirow{4}{4em}{AUC} & $A$ & 0.846(7) & 0.809(7) & \textbf{0.848(5)} & 0.837(8) \\
     & $h^0$ & \textbf{0.752(5)} & 0.742(5) & 0.701(5) & 0.728(4) \\
     & $h^\pm$ & 0.914(3) & \textbf{0.917(4)} & 0.817(11) & 0.912(3) \\
     & $LQ$ & \textbf{0.851(4)} & 0.840(6) & 0.770(12) & 0.835(6) \\
    \hline
    \multirow{4}{4em}{$\epsilon_b^{-1}$ $(\epsilon_s = 0.3)$} & $A$ & 18.0(8) & 12.9(8) & \textbf{48.1(116)} & 17.7(10) \\
     & $h^0$ & 18.0(16) & 17.0(27) & 10.8(8) & \textbf{18.2(17)} \\
     & $h^\pm$ & \textbf{76.5(192)} & 62.8(108) & 22.3(30) & 75.7(197) \\
     & $LQ$ & \textbf{19.9(2)} & 16.3(10) & 13.7(7) & 19.8(3) \\
    \hline
    \multirow{4}{4em}{SI $(\epsilon_s = 0.3)$} & $A$ & 1.27(3) & 1.07(3) & \textbf{2.06(23)} & 1.26(3) \\
     & $h^0$ & 1.27(6) & 1.23(9) & 0.984(35) & \textbf{1.28(6)} \\
     & $h^\pm$ & \textbf{2.59(31)} & 2.36(20) & 1.41(9) & 2.58(32) \\
     & $LQ$ & \textbf{1.34(1)} & 1.21(4) & 1.11(3) & 1.33(1) \\
    \hline
    \multirow{4}{4em}{Max F1 $(\epsilon_s)$} & $A$ & \textbf{0.795(9)} $(\epsilon_s = 0.85)$ & 0.774(7) $(\epsilon_s = 0.89)$ & 0.781(8) $(\epsilon_s = 0.83)$ & 0.794(8) $(\epsilon_s = 0.85)$ \\
     & $h^0$ & \textbf{0.717(4)} $(\epsilon_s = 0.93)$ & 0.714(4) $(\epsilon_s = 0.93)$ & 0.686(8) $(\epsilon_s = 0.90)$ & 0.693(3) $(\epsilon_s = 0.81)$ \\
     & $h^\pm$ & 0.846(3) $(\epsilon_s = 0.88)$ & \textbf{0.854(4)} $(\epsilon_s = 0.88)$ & 0.764(11) $(\epsilon_s = 0.86)$ & 0.851(4) $(\epsilon_s = 0.90)$ \\
     & $LQ$ & \textbf{0.796(3)} $(\epsilon_s = 0.90)$ & 0.790(5) $(\epsilon_s = 0.87)$ & 0.732(11) $(\epsilon_s = 0.84)$ & 0.788(7) $(\epsilon_s = 0.86)$ \\
    \hline
    \multirow{4}{4em}{Max SI $(\epsilon_s)$} & $A$ & 1.73(4) $(\epsilon_s = 0.68)$ & 1.46(3) $(\epsilon_s = 0.77)$ & \textbf{2.06(27)} $(\epsilon_s = 0.28)$ & 1.70(5) $(\epsilon_s = 0.67)$ \\
     & $h^0$ & 1.29(8) $(\epsilon_s = 0.24)$ & 1.25(7) $(\epsilon_s = 0.39)$ & 1.09(2) $(\epsilon_s = 0.65)$ & \textbf{1.29(7)} $(\epsilon_s = 0.24)$ \\
     & $h^\pm$ & 2.63(26) $(\epsilon_s = 0.39)$ & 2.53(17) $(\epsilon_s = 0.58)$ & 1.50(6) $(\epsilon_s = 0.60)$ & \textbf{2.65(27)} $(\epsilon_s = 0.39)$ \\
     & $LQ$ & 1.66(3) $(\epsilon_s = 0.72)$ & 1.60(4) $(\epsilon_s = 0.69)$ & 1.29(5) $(\epsilon_s = 0.69)$ & \textbf{1.66(4)} $(\epsilon_s = 0.63)$ \\
    \hline
    \end{tabular}}
    \caption{
    Anomaly-detection performance of the OT-only ($N=2,000$ reference events) and OT+ML methods for the 3D arc ground space.
    The performance metrics are the area under the ROC curve (AUC), the inverse background efficiency ($\epsilon_b^{-1}$), and the significance improvement (SI), where the latter two are quoted at $\epsilon_s = 0.3$.
    We also report the maximum F1 and SI values, together with the signal efficiency at which each maximum occurs; the maximum SI is taken over $\epsilon_s \in [0.1, 1]$.
    Each entry is quoted as mean(error): the mean is taken over five independent sets of test events, and the digits in parentheses give the corresponding standard deviation in units of the last digit of the mean. For example, $0.846(7)$ denotes $0.846 \pm 0.007$ and $76.5(192)$ denotes $76.5 \pm 19.2$.
    Bold entries mark the best-performing method for each metric and signal.
    Overall, 3D minOT performs best, although 3D OT + one-class SVM follows closely.
    3D OT + anomaly kNN is the weakest method for every signal except $A$, where it generally gave the best performance. However, 3D minOT's AUC is slightly lower but within the uncertainty range.
    }
    \label{tab:OT_anomaly_detection_comparison}
\end{table}

Given the multitude of combinations of ground spaces and methods, we begin by comparing the AUC of the closest OT methods (i.e. minOT and QminOT) to help indicate the best performing ground space.
Figure~\ref{fig:OTonly_AUC_comparison} compares the AUC values for the closest OT methods under different ground space choices across all four signals.
We can see that the 3D distances outperform the 2D distances for all closest OT methods, making them the best overall ground space choice. 
However, the 2D total $p_{\rm T}$ difference recovers a significant amount of the performance gap between the 2D arc distance and the 3D distances.
Despite this boost, a gap still remains.
This indicates that total $p_{\rm T}$ difference is important but the 3D distances are also capturing information about the relative distributions of $p_{\rm T}$ information.
We also see that 3D arc distances and 3D chord distances perform equivalently well.
Additionally, we can see that the performance of minOT and $\text{QminOT}_p$ are approximately equivalent, and in fact are identical for small enough $p$, as expected. 
Differences in the stability of minOT and QminOT with increasing $N$ are explored in Appendix~\ref{app:scaling_with_N}.

Having found that 3D distances are the best ground space choice, we fix our choice of 3D arc distances when comparing closest OT and OT+ML methods.
Figure~\ref{fig:ROC_curves} plots the mean and $\pm 1$-standard deviation error band of the ROC curves for 3D closest OT and 3D OT+ML methods across all signals.
Figure~\ref{fig:SI_curves} plots the mean and $\pm 1$-standard deviation error band of the SI curves for 3D closest OT and 3D OT+ML methods across all signals. Note that a lower signal efficiency threshold ($\epsilon_s=0.10$) is imposed since the SI curve becomes unstable for low signal efficiencies.

Figures~\ref{fig:ROC_curves} and~\ref{fig:SI_curves} show that 3D OT + anomaly kNN is the weakest of the four methods for most signals ($h^0$, $h^\pm$, and $LQ$).
We attribute this largely to the fact that in order to make it a weakly supervised anomaly detection, anomaly augmented background events were used in place of a signal model to train the kNN algorithm. 
However, these anomaly augmented background events do not fully capture the structure of the signal models themselves.
The comparison with supervised event classification, which couples the same 3D OT distances to the kNN algorithm (Table~\ref{tab:kNNOT_classification} in Appendix~\ref{app:classification}), makes this explicit: the substantially better classification performance indicates how much is lost by training the kNN on anomaly-augmented backgrounds rather than on genuine signal.
The signal $A$ case is the exception, here the 3D OT + anomaly kNN method outperforms the other methods, especially at low signal efficiency.
Unlike the other signal models, for $A$ the total multiplicity is a much stronger baseline discriminant than the total $p_{\rm T}$, and two of the three anomaly augmentations used to construct the anomaly-augmented background dataset --- which subsequently serves as the signal proxy when training 3D OT + anomaly kNN --- capture precisely this property.
The augmented backgrounds are therefore a closer proxy for genuine $A$ events, which is reflected in the smaller gap between the anomaly-detection and classification results for this signal. 
This largely explains why 3D OT + anomaly kNN slightly outperforms the direct use of 3D OT distances.

The remaining three methods, 3D minOT, 3D QminOT$_{1\%}$, and 3D OT + one-class SVM, perform comparably across all signals, with the one-class SVM closely tracing the minOT ROC curve in every case.
This can be understood by viewing minOT and QminOT as being similar to the large-$\gamma$ limit of the one-class SVM.
One-class SVMs create a decision boundary and a test event's anomaly score is the closest distance to this boundary. 
In the limit of large $\gamma$ every event contributes to this decision boundary making it wrap tightly around each reference event. Thus, a one-class SVM's score becomes (approximately) the distance to the closest event in this limit.
While the $\gamma$ preferred in our parameter scans is comparatively large, we still see the closest OT methods win out at very high signal efficiency values for both the $h^0$ and ${\rm LQ}$ signal cases.

Overall, the closest OT methods perform the best especially at very high $\epsilon_s$.
This is intriguing as they are simple and inherently unsupervised methods which already approach the performance of state-of-the-art ML methods combining autoencoders with contrastive learning~\cite{Dillon:2023zac} (see  Table~\ref{tab:baseline_performance} in Appendix~\ref{app:baseline_performance}). 
While the performance of model-dependent baselines is already quite high ( Table~\ref{tab:baseline_performance}), the closest-OT methods' performance are comparable to the best baseline method for each signal model, making it an all-in-one anomaly score.

In principle, closest OT methods may be used for anomaly detection both at trigger level and in offline analyses. 
There are two main considerations governing the feasibility of using closest OT methods in an online setting for L1 trigger-level data: memory limitations and computational evaluation time~\cite{Govorkova:2021hqu}. 
State-of-the-art ML methods are typically fast to evaluate but may require significant pruning to meet memory constraints~\cite{Belis:2023mqs}.
In general, calculating the OT distance between two events is computationally costly, but many approximation strategies can lower this computational burden~\cite{cuturi2013sinkhorndistanceslightspeedcomputation, altschuler2018, sliced_wasserstein, kolouri2019generalizedslicedwassersteindistances, nguyen2023, Cai:2020vzx}.
Moreover, the computational cost is less intense when the number of probability masses is low (which is the case for this L1 trigger dataset), and scales very well with an increase in the dimension of the underlying ground space.
The memory limitations for OT methods are also likely to be less intense than state-of-the-art ML methods.  
For OT methods, the main memory requirement would be storing a representative sample of SM background events to compare a test event to.  
Both of these considerations (memory and computational cost) can be lessened by calculating the OT distances between a test event and the $k$-medoids of the SM background data, as was done in previous work~\cite{Fraser:2021lxm, CrispimRomao:2020ejk, Komiske:2019fks}. 
Alternatively, the costs may be significantly reduced by using linearized OT \cite{Cai:2020vzx}, which provides an isometric linear embedding of all the background events into Euclidean space. 
Using this linearized strategy, determining the anomaly score for an event using the closest 3D OT distance to the background sample would then be reduced to computing a single 3D OT distance to a suitably-chosen reference event, combined with the efficient calculation of Euclidean distances to events in the background sample. 
We discuss the computational efficiency of the proof-of-concept results shown in this paper as well as what must be done for it to be feasibly deployed at the trigger level in Appendix~\ref{app:ot_timing}.
However, a comprehensive study investigating these considerations would be beneficial and is left to future work.

\section{Conclusion}\label{sec:conclusion}

The plethora of possible extensions of the Standard Model (and the correspondingly vast space of potential signals) motivates the development of model-agnostic search techniques such as anomaly detection. 
In this paper, we have initiated the use of balanced OT distances for event-level anomaly detection, both on their own and coupled to interpretable, unsupervised (or weakly supervised) machine learning algorithms. 
In both cases, the 2-Wasserstein distance based on a three-dimensional ground space that includes $p_T/{\rm GeV}$ as a coordinate alongside the physical $\eta, \phi$ calorimeter coordinates (``3D OT distances'') outperforms the 2-Wasserstein distance using the $\eta, \phi$ coordinates alone (``2D OT distances''). 
We find that two versions of the 3D coordinates ($3\rm D_{\rm arc}$ and $3\rm D_{\rm chord}$) perform equivalently. 
We also find that some, but not all, of the performance gap between 2D and 3D can be closed by including information about the $p_{\rm T}$ difference between events ($\rm2D_{\Delta p_{\rm T}}$).
This means there is still interesting information about the distribution of $p_{\rm T}$ within an event which is useful for identifying BSM anomalies with OT distances.
Given its promising performance we primarily focused on $3\rm D_{\rm arc}$ distances in the majority of the paper.

Using these distances we considered two strategies for developing anomaly scores which both capitalize on the inherent feature that OT distances will be smaller for like-distribution events.
The first, used the closest-distance to an ensemble of $N$ SM events as an anomaly score, where ``closest'' meant either the minimum distance (${\rm minOT}$) or the $p\%$ quantile (${\rm QminOT_{p\%}}$) where $p\geq 1/N$. In the case that $p=1/N$ these two methods are equivalent.
Interestingly, we found that a small $p$ was often preferred and that the 3D${\rm ~minOT}$ often had the best performance.
While empirically both strategies were relatively stable with increasing $N$  (see Appendix~\ref{app:scaling_with_N}), ${\rm QminOT_{p\%}}$ is theoretically more stable to fluctuations in the reference set.

The second strategy focuses on coupling OT distances to simple, distance-based machine learning algorithms such as kNN (3D OT+anomaly kNN) or one-class SVM (3D OT+one-class SVM). 
Overall, we found that the closest distance strategies achieved better performance than the OT+ML strategies, particularly for
the 3D OT+anomaly kNN. 
In the 3D OT+one-class SVM case, the closest-distance approaches can be seen as a large $\gamma$ limit of a one-class SVM. 
Thus, predictably the performance tracked more closely but still underperformed in comparison, particularly at high signal efficiency.

Although this simple closest OT distance approach to anomaly detection does not fully match the performance of more sophisticated ML methods on the same dataset~\cite{Dillon:2023zac}, it demonstrates the key features and potential advantages of using optimal transport for anomaly detection in collider data.
Computational improvements will be necessary to scale this approach to the level necessary for L1 trigger deployment. 
However, many efficient approximation strategies may be useful in this pursuit. We discuss these in more detail in
Appendix~\ref{app:ot_timing} but leave a detailed comparison study to future work.

To improve the performance of these baselines, there are two main directions for future work. 
The first is to incorporate more information into the optimization problem.
For example, one could consider adding additional ground space dimensions, e.g. multiplicity, to capitalize on known differences between SM and many BSM signals.
Along these lines, one could consider encoding species-type information, i.e. multi-species OT. 
Ref.~\cite{CrispimRomao:2020ejk} investigated adding species difference as a weight on the ground space distance.  There are a number of well-posed forms of multi-species optimal transport that may prove fruitful in this setting~\cite{Cai:2021hnn}. 

The second strategy for future work is to improve the correspondence between physical differences and OT distances between events.
Concretely, one could achieve this by accounting for symmetry transformations of the underlying ground space. Ref.~\cite{Larkoski:2023qnv} investigated choosing a ground space which is manifestly invariant to event-level physical transformations.
The $3\rm D_{\rm chord}$ distance considered here already makes strides towards this by handling azimuthal symmetry, but more general strategies are needed.
Another straightforward, and mathematically rigorous, way to handle this would be to add ground-space transformations (e.g. rotations and reflections) to the problem as additional optimization steps~\cite{toappear}. 
It might also be interesting to consider other varieties of OT distances which are manifestly insensitive to such ground-space transformations. 
Namely, the Gromov-Wasserstein distance~\cite{memoli2011gromov}, originally designed to compute OT distances between distributions with ground spaces of different dimensions, only depends on the relative locations of the probability masses within each event.
It is therefore, manifestly invariant to translations and rotations of the underlying ground space.
While the Gromov-Wasserstein distance itself is computationally non-trivial, it admits several approximations~\cite{li2023efficient} which are computationally efficient for discrete distributions with a small number of probability masses.

\section*{Acknowledgements}
We would like to thank Katy Craig and Tianji Cai for useful conversations about optimal transport, and Tianji Cai for sharing her kNN classification code.
We would also like to thank Gregor Kasieczka, Cristiano Fanelli, Thomas G. Rizzo, and Michael Peskin for useful discussions.
This work was performed in part at the Aspen Center for Physics, which is supported by National Science Foundation grant PHY-2210452.
JNH was supported by the National Science Foundation under Grant No. NSF PHY-1748958, by the Gordon and Betty Moore Foundation through Grant No. GBMF7392, and by a UCSB Chancellor's postdoctoral fellowship. 
NC was supported by the Department of Energy under Grant No.~DE-SC0011702. 

\begin{appendix}

\section{Appendix: Baseline performance} \label{app:baseline_performance}
In this appendix, we summarize the performance of common baseline methods. When evaluating these baseline methods (Total $p_{\rm T}$, MET, and Total Multiplicity) we construct a test set from $1,000$ SM background events and $1,000$ BSM signal events (for each signal model). 
For comparison, we also include the results from a recent work studying the performance of various autoencoder methods on this dataset~\cite{Dillon:2023zac}. 

\begin{table}[ht]
    \centering
    \resizebox{\textwidth}{!}{%
    \begin{tabular}{c c c|c|c|c|c c|c|}
    && & \multicolumn{3}{c|}{Common Baselines} && \multicolumn{2}{c|}{Recent Results from Ref.~\cite{Dillon:2023zac}} \\
    && & Total $p_{\rm T}$ & ${\rm MET}$ & Total Multiplicity && AE-Raw & AnomCLR+ \\
    \hline
    \multirow{4}{4em}{AUC}
       && $A$     & 0.741(11) & 0.369(14) & \textbf{0.891(6)} && 0.885(2) & \textbf{0.909(3)} \\
       && $h^0$   & 0.708(9)  & 0.598(15) & \textbf{0.738(9)} && 0.755(2) & \textbf{0.776(2)} \\
       && $h^\pm$ & \textbf{0.926(4)} & 0.837(9) & 0.878(5)     && 0.900(4) & \textbf{0.930(1)} \\
       && $LQ$    & 0.842(7)  & 0.561(11) & \textbf{0.875(7)} && 0.856(2) & \textbf{0.880(1)} \\
    \hline
    \multirow{4}{4em}{$\epsilon_b^{-1}$ $(\epsilon_s = 0.3)$}
       && $A$     & 13.3(15)  & 2.01(11)  & \textbf{82.2(171)} && 47(2)     & \textbf{139(20)} \\
       && $h^0$   & \textbf{20.0(37)} & 8.26(108) & 13.6(14)   && 14.9(7)   & \textbf{23(1)} \\
       && $h^\pm$ & 88.3(171) & \textbf{389(335)} & 46.9(58)   && 60(10)    & \textbf{171(7)} \\
       && $LQ$    & 19.9(13)  & 4.78(38)  & \textbf{38.1(52)}  && 24.4(60)  & \textbf{39(1)} \\
    \hline
    \multirow{4}{4em}{SI $(\epsilon_s = 0.3)$}
       && $A$     & 1.09(6)   & 0.425(11) & \textbf{2.69(27)} && 2.05(5)   & \textbf{3.5(2)} \\
       && $h^0$   & \textbf{1.33(12)} & 0.860(55) & 1.10(6)   && 1.16(3)   & \textbf{1.44(3)} \\
       && $h^\pm$ & 2.79(27) & \textbf{5.28(221)} & 2.04(12)  && 2.3(2)    & \textbf{3.9(1)} \\
       && $LQ$    & 1.33(4)   & 0.655(26) & \textbf{1.84(12)} && 1.48(2)   & \textbf{1.88(2)} \\
    \hline
    \multirow{4}{4em}{Max F1 $(\epsilon_s)$}
       && $A$     & 0.714(3) $(\epsilon_s = 0.86)$ & 0.667(0) $(\epsilon_s = 1.00)$ & \textbf{0.825(7)} $(\epsilon_s = 0.83)$ && - & - \\
       && $h^0$   & 0.671(8) $(\epsilon_s = 0.77)$ & 0.667(0) $(\epsilon_s = 1.00)$ & \textbf{0.721(3)} $(\epsilon_s = 0.91)$ && - & - \\
       && $h^\pm$ & \textbf{0.864(4)} $(\epsilon_s = 0.89)$ & 0.754(7) $(\epsilon_s = 0.76)$ & 0.806(7) $(\epsilon_s = 0.91)$ && - & - \\
       && $LQ$    & 0.793(8) $(\epsilon_s = 0.90)$ & 0.667(0) $(\epsilon_s = 1.00)$ & \textbf{0.812(8)} $(\epsilon_s = 0.93)$ && - & - \\
    \hline
    \multirow{4}{4em}{Max SI $(\epsilon_s)$}
       && $A$     & 1.20(4) $(\epsilon_s = 0.68)$  & 1.00(0) $(\epsilon_s = 1.00)$ & \textbf{2.76(17)} $(\epsilon_s = 0.42)$ && - & - \\
       && $h^0$   & \textbf{1.40(23)} $(\epsilon_s = 0.20)$ & 1.05(14) $(\epsilon_s = 0.20)$ & 1.17(3) $(\epsilon_s = 0.72)$ && - & - \\
       && $h^\pm$ & 3.13(64) $(\epsilon_s = 0.20)$ & \textbf{8.16(700)} $(\epsilon_s = 0.22)$ & 2.09(10) $(\epsilon_s = 0.54)$ && - & - \\
       && $LQ$    & 1.61(6) $(\epsilon_s = 0.65)$  & 1.00(0) $(\epsilon_s = 1.00)$ & \textbf{1.95(7)} $(\epsilon_s = 0.51)$  && - & - \\
    \hline
    \end{tabular}}
    \caption{
    Performance of the baseline methods for each signal dataset. The performance metrics considered are the area under the ROC curve (AUC), the inverse background efficiency ($\epsilon_b^{-1}$), and the significance improvement (SI), where the latter two are reported at a signal efficiency of $\epsilon_s = 0.3$. We also report the maximum F1 and SI values together with the signal efficiency at which each maximum occurs, taken over the range $\epsilon_s \in [0.2, 1]$. The lower threshold $\epsilon_s = 0.2$ is imposed when locating the maximum SI because the SI curve becomes unstable at low signal efficiencies. Each entry is quoted as mean(error), where the digits in parentheses give the standard deviation in units of the last digit of the mean, so that $0.741(11)$ denotes $0.741 \pm 0.011$ and $139(20)$ denotes $139 \pm 20$. Bold entries mark the best-performing method within each group, and a dash indicates a quantity not reported in Ref.~\cite{Dillon:2023zac}.
    We only consider baseline methods which do not make reference to the identities of particles (e.g.\ lepton multiplicity). Even though these methods are standard~\cite{Mikuni:2021nwn}, comparison to such quantities would be misleading since the OT methods employed do not currently have access to particle identity information. For comparison, we also include the results from a recent work studying the performance of various autoencoder methods on this dataset~\cite{Dillon:2023zac}.
    }
    \label{tab:baseline_performance}
\end{table}

\section{Appendix: Comparison with Classification Performance} \label{app:classification}

\begin{table}[ht]
    \centering
    \resizebox{\textwidth}{!}{%
    \begin{tabular}{c c|c|c|c c|c|c|}
    && \multicolumn{6}{c|}{OT+kNN classification performance} \\
    && 2D & 2D$_{\Delta p_T}$ & \multicolumn{2}{c|}{2D planed total $p_{\rm T}$ (GeV)} & 3D$_{\rm arc}$ & 3D$_{\rm chord}$ \\
    && & & $p_{\rm T} \in (50,100)$ & $p_{\rm T} \in (500,1000)$ & & \\
    \hline
    \multirow{4}{4em}{AUC} & $A$ & 0.738(5) & 0.777(5) & 0.797(9) & 0.799(22) & 0.913(7) & \textbf{0.913(7)} \\
     & $h^0$ & 0.682(11) & 0.740(7) & 0.676(36) & 0.583(21) & 0.775(7) & \textbf{0.777(7)} \\
     & $h^\pm$ & 0.813(8) & 0.912(3) & 0.607(27) & 0.620(34) & 0.924(3) & \textbf{0.924(4)} \\
     & $LQ$ & 0.797(7) & 0.837(4) & 0.747(21) & 0.528(15) & \textbf{0.879(4)} & 0.879(4) \\
    \hline
    \multirow{4}{4em}{$\epsilon_b^{-1}$ $(\epsilon_s = 0.3)$} & $A$ & 11.0(3) & 13.3(8) & 32.5(68) & 27.2(140) & \textbf{150(73)} & 142(73) \\
     & $h^0$ & 8.20(79) & 17.6(29) & 7.73(214) & 4.96(61) & 19.8(35) & \textbf{19.9(38)} \\
     & $h^\pm$ & 20.4(26) & 72.2(117) & 5.75(154) & 6.30(155) & 75.6(122) & \textbf{77.3(106)} \\
     & $LQ$ & 16.8(15) & 20.0(23) & 11.4(30) & 3.86(58) & 36.5(49) & \textbf{38.0(66)} \\
    \hline
    \multirow{4}{4em}{SI $(\epsilon_s = 0.3)$} & $A$ & 0.993(12) & 1.09(3) & 1.70(17) & 1.51(39) & \textbf{3.56(77)} & 3.45(79) \\
     & $h^0$ & 0.858(41) & 1.25(11) & 0.827(108) & 0.666(41) & 1.33(11) & \textbf{1.33(13)} \\
     & $h^\pm$ & 1.35(8) & 2.53(20) & 0.714(89) & 0.747(93) & 2.59(21) & \textbf{2.62(18)} \\
     & $LQ$ & 1.23(5) & 1.34(8) & 1.00(13) & 0.587(45) & 1.81(12) & \textbf{1.84(15)} \\
    \hline
    \multirow{4}{4em}{Max F1 $(\epsilon_s)$} & $A$ & 0.718(5) $(\epsilon_s = 0.87)$ & 0.743(4) $(\epsilon_s = 0.88)$ & 0.749(15) $(\epsilon_s = 0.86)$ & 0.743(18) $(\epsilon_s = 0.77)$ & 0.839(8) $(\epsilon_s = 0.85)$ & \textbf{0.841(7)} $(\epsilon_s = 0.86)$ \\
     & $h^0$ & 0.687(6) $(\epsilon_s = 0.89)$ & 0.707(2) $(\epsilon_s = 0.89)$ & 0.688(18) $(\epsilon_s = 0.89)$ & 0.669(11) $(\epsilon_s = 0.99)$ & 0.734(4) $(\epsilon_s = 0.91)$ & \textbf{0.734(4)} $(\epsilon_s = 0.91)$ \\
     & $h^\pm$ & 0.763(4) $(\epsilon_s = 0.86)$ & 0.846(2) $(\epsilon_s = 0.87)$ & 0.668(16) $(\epsilon_s = 0.98)$ & 0.673(19) $(\epsilon_s = 0.92)$ & \textbf{0.860(3)} $(\epsilon_s = 0.90)$ & 0.860(1) $(\epsilon_s = 0.89)$ \\
     & $LQ$ & 0.757(7) $(\epsilon_s = 0.88)$ & 0.784(6) $(\epsilon_s = 0.88)$ & 0.723(17) $(\epsilon_s = 0.83)$ & 0.667(16) $(\epsilon_s = 0.99)$ & \textbf{0.814(2)} $(\epsilon_s = 0.90)$ & 0.814(2) $(\epsilon_s = 0.90)$ \\
    \hline
    \multirow{4}{4em}{Max SI $(\epsilon_s)$} & $A$ & 1.18(2) $(\epsilon_s = 0.71)$ & 1.30(1) $(\epsilon_s = 0.66)$ & 1.88(47) $(\epsilon_s = 0.25)$ & 1.71(45) $(\epsilon_s = 0.40)$ & 3.62(81) $(\epsilon_s = 0.27)$ & \textbf{3.67(176)} $(\epsilon_s = 0.16)$ \\
     & $h^0$ & 1.07(2) $(\epsilon_s = 0.78)$ & 1.28(11) $(\epsilon_s = 0.20)$ & 1.09(9) $(\epsilon_s = 0.69)$ & 1.00(1) $(\epsilon_s = 0.99)$ & 1.36(14) $(\epsilon_s = 0.21)$ & \textbf{1.38(20)} $(\epsilon_s = 0.18)$ \\
     & $h^\pm$ & 1.46(3) $(\epsilon_s = 0.62)$ & 2.76(17) $(\epsilon_s = 0.48)$ & 1.00(1) $(\epsilon_s = 0.98)$ & 1.02(3) $(\epsilon_s = 0.90)$ & 2.87(24) $(\epsilon_s = 0.40)$ & \textbf{2.92(26)} $(\epsilon_s = 0.44)$ \\
     & $LQ$ & 1.36(3) $(\epsilon_s = 0.73)$ & 1.61(3) $(\epsilon_s = 0.64)$ & 1.26(5) $(\epsilon_s = 0.67)$ & 1.00(1) $(\epsilon_s = 0.99)$ & \textbf{1.94(8)} $(\epsilon_s = 0.51)$ & 1.92(9) $(\epsilon_s = 0.53)$ \\
    \hline
    \end{tabular}}
    \caption{
    Classification performance of the kNN+OT method for background against each signal model using $N=2000$ events per class. The performance metrics considered are the area under the ROC curve (AUC), inverse background efficiency rate ($\epsilon_b^{-1}$), and significance improvement (SI), where the latter two are reported at a signal efficiency rate of $\epsilon_s = 0.3$. We also report the maximum SI and F1 values and their corresponding $\epsilon_s$ values. For the non-planed columns, the maximum SI is computed over $\epsilon_s \in [0.1, 1]$. The 2D planed total $p_{\rm T}$ columns give the low and high total-$p_{\rm T}$ ranges considered. Each entry is quoted as mean(error), where the mean is taken over five independent repeats and the digits in parentheses give the corresponding standard deviation in units of the last digit of the mean, so that $0.738(5)$ denotes $0.738 \pm 0.005$ and $150(73)$ denotes $150 \pm 73$. Bold entries mark the best-performing ground space for each metric and signal. The optimal hyperparameters are stored in the \href{https://github.com/hancheng-li/OT_anomaly_detection}{GitHub repo}.
    This supervised classification performance gives a best-case performance limit for the OT+kNN anomaly detection strategy. We can see that 3D OT distances contain more discerning information, far outperforming their 2D counterparts. We also see that ${\rm 3D_{arc}}$ vs ${\rm 3D_{chord}}$ distances perform equivalently. 
    Additionally, we see that including total $p_{\rm T}$ differences significantly improves the 2D distance. 
    }
    \label{tab:kNNOT_classification}
\end{table}

To investigate the information contained in the 2D vs 3D ground spaces, we looked at the performance of a supervised kNN algorithm trained to distinguish SM from each BSM model individually. 
This serves as a best-case performance limit for the kNN anomaly detection algorithm.
Indeed we find that the 3D ground space contains much more information than its 2D counterpart. 
This is likely due to the importance of total transverse momentum, $p_{\rm T}$, in this dataset (see Table~\ref{tab:baseline_performance}).
To test this, we also consider kNN algorithms trained on events with ``planed'' total $p_{\rm T}$~\cite{Chang:2017kvc}. 
We find that, in the $A$ case, the 2D ground space achieves a performance boost both compared to the un-planed case as well as the total $p_{\rm T}$.
This indicates that, for some signal models, there can be extra discriminating information contained in the $\eta,\phi$ distribution of an event.
We also note, that for all signal cases, the 3D OT+kNN classification performance outperforms the 3D OT+kNN anomaly detection performance.  
This indicates, that there is still more discriminating information that could be extracted.

\section{Appendix: Memory and Computational Requirements of OT Calculation} \label{app:ot_timing}
In this proof of principle work we focused on using the standard OT distance implementation provided by the {\sc POT} library~\cite{flamary2021pot} following the algorithm of Ref.~\cite{Bonneel2011}. 
This was done to minimize additional uncertainties introduced from more efficient approximation strategies.

The computational cost of an OT calculation is governed primarily by the number of discrete probability masses, $n$, describing each event.
Fortunately, this is fixed and comparatively small in our case ($n=19$). Thus, what really matters in this context is the size of the coefficient. 
Let $C$ be the cost of a single OT distance between two events, ${\rm OT}(\mathcal{E}, \mathcal{E}')$. 
In our experiments, the average runtime required to evaluate a single standard OT distance between two events, ${\rm OT}(\mathcal{E}, \mathcal{E}')$, on a standard laptop is $\mathcal{O}(10)$ ${\rm \mu s}$ (see Table~\ref{tab:standardOTtiming}). 
This is unfortunately too slow to be ready for real-time deployment, where the time interval between events is $25~{\rm ns}$ and the entire L1 trigger algorithm must deploy in $\mathcal{O}(1)$ $\mu s$~\cite{Govorkova:2021hqu}.
Since the calculation time must be reduced by at least 3 orders of magnitude, approximation strategies are likely required. 

\begin{table}[ht]
    \centering
    \begin{tabular}{|c|c|c|c|}
    \hline
    \multicolumn{4}{|c|}{Standard OT Algorithm Computation Time ($\mu {\rm s}$)} \\
    \hline
    2D & 2D$_{\Delta p_{\rm T}}$ & 3D$_{\rm arc}$ & 3D$_{\rm chord}$\\
    \hline
    $95.6 \pm 14.3$ & $52.9 \pm 7.95$ & $46.6 \pm 8.19$ & $63.9 \pm 10.1$ \\
    \hline
    \end{tabular}
    \caption{Computational time required to calculate the OT distance between two events using the standard implementation provided by the POT library~\cite{flamary2021pot} following the algorithm of Ref.~\cite{Bonneel2011}. These times were averaged over 1,000 computations on a MacBook Pro with an Apple M3 Pro chip and 18GB of memory. We report the mean and standard deviation. While currently too slow for online deployment many promising approximation strategies exist.}
    \label{tab:standardOTtiming}
\end{table}

Below, we summarize the range of approximation strategies that exist and what this implies about a path towards a practical implementation of trigger-level OT anomaly detection where both memory and computational time are limited. 
For simplicity we focus on a serialized estimation of computational cost, but note that many aspects of these calculations are amenable to parallelization.
We will first focus on the requirements of the distance-only strategies (minOT and QminOT) and then comment on the requirements of the OT + oneClassSVM strategy.

\subsection{OT distance approximations}
There are three main OT approximation strategies which can reduce the computational cost: 
(1) Sinkhorn distances~\cite{cuturi2013sinkhorndistanceslightspeedcomputation, altschuler2018}, 
(2) Sliced Wasserstein distances~\cite{sliced_wasserstein, kolouri2019generalizedslicedwassersteindistances,nguyen2023}, 
(3) Linearized OT (LOT) distances\cite{Wang2013LinearOT, Cai:2020vzx}. 
Methods (1) and (2) will simply reduce $C$ at the risk of potentially making the distance estimates noisier. 
While these are the preferred choices for calculating OT distances at-scale, the typical computational benefits center on their improved scaling with $n$. 
The standard strategy has $\mathcal{O}(n^3)$ scaling, whereas method (2) reduces this to $\mathcal{O}(n \log n)$~\cite{nguyen2023}  and method (1) to near linear performance $\mathcal{O}(n)$~\cite{altschuler2018}.%
\footnote{ 
We caution that since we are operating on events with small $n$, implementation differences matter greatly. 
A fair comparison may require more specialized implementations of these algorithms than what exists in the POT library.
This is because the use of backend architecture may introduce overhead costs that are not representative of an algorithm's actual performance. 
We leave a detailed study of algorithm selection and feasibility of an implementation on FPGAs to future work.
}
Thus these strategies pay off the most for very large $n$.

On the other hand, method (3) reduces the number of OT calculations one has to perform. 
In particular, assume we have to calculate $N_{\rm ref}$ OT distances between a test event $\mathcal{T}$ and our set of $N_{\rm ref}$ reference background events $\{\mathcal{S}_1, \cdots, \mathcal{S}_{N_{\rm ref}}\}$. The cost of doing this would be $\mathcal{O}\left(C N_{\rm ref}\right)$. Instead, LOT exploits the local geometric structure of the 2-Wasserstein distance space by introducing a tangent-space representation around
a fixed reference event, $\mu_0$, comprising $n'$ probability point masses.%
\footnote{In practice, $\mu_0$ is typically chosen to be a synthetic distribution that spans the ground space, such as a regular grid of $n'$ point masses. Where $n'$ need not equal $n=19$ but typically will be chosen to be of the same order or larger.}
Each reference event $\mathcal{S}$ is embedded into this tangent space by solving a single OT problem between the event and $\mu_0$, producing a transport plan, $\gamma \in \mathbb{R}^{n' \times n}$ that serves as its representation. 
Similarly, we obtain the transport plan from the OT problem between our test and reference events, $\gamma_T \in \mathbb{R}^{n' \times n}$.
The LOT distance~\cite{Cai:2020vzx} between our test event and a reference event $\mathcal{S}$ is then given by \begin{align}
    {\rm LOT}(\mathcal{S}, \mathcal{T}) = \left(n' \sum_{i=1}^{n'} \Vert z_i - \tilde{z}_i\Vert^2\right)^{1/2} 
\end{align}
where $z_i$ ($\tilde{z}_i$) is the $i^{\rm th}$ row of the $n' \times d$ matrix $z = \gamma \cdot x$ ($\tilde{z} = \gamma_T \cdot \tilde{x}$) and $x$ ($\tilde{x}$) is a $n \times d$ matrix denoting the locations of the point probability masses in the reference (test) event.
Therefore, this has an overall computational scaling of $\mathcal{O}(n' n)$.%
\footnote{We neglect the scaling with $d$ since it will be much less than $n,n'$.}
Thus, the computational cost would be $\mathcal{O}\left(C + N_{\rm ref} n' n \right)$.
While LOT will likely give a more faithful approximation this strategy will also increase the memory footprint as the reference transport plans must be stored.
Additionally, since the transport plan is required, distance-only approximation strategies like sliced wasserstein distances cannot be used.

\subsection{Computational scaling of current algorithms}
For the closest-distance algorithms, scoring a single event $\mathcal{T}$ requires $N_{\rm ref}$ OT computations in addition to an $O(1)$ sorting algorithm.
Thus, the computational cost would be $\mathcal{O}\left((C+1) N_{\rm ref}\right)$. However, the memory footprint is quite small as only $N_{\rm ref}$ events must be stored and each $(19,3)$ event is $57$ floating point numbers.  
For OT + One Class SVM, the OT distance computation requirements are the same but instead of a $\mathcal{O}(1)$ algorithm, we would have the cost of the model's evaluation, $C_{\rm eval}$. Thus the computational cost would be  $\mathcal{O}\left(C N_{\rm ref} + C_{\rm eval} \right)$.
The memory footprint in this case would just be the number of support vectors of the model times the size of those events, $N_{SV} \times 57$ floating point numbers.

If LOT distances are used as the OT distance approximation strategy, the computational scaling would be $\mathcal{O}\left(C + N_{\rm ref} n' n \right)$, but the memory footprint would increase substantially.
Given that we must store $N_{\rm ref}$ $n' \times n$ transport plans in addition to the reference event $\mu_0$. This means we would need to store $ nn'N_{\rm ref} + 57$ floating point numbers.

\section{Appendix: How minOT and QminOT scale with $N$} \label{app:scaling_with_N}
\begin{figure}
    \centering
    \includegraphics[width=\linewidth]{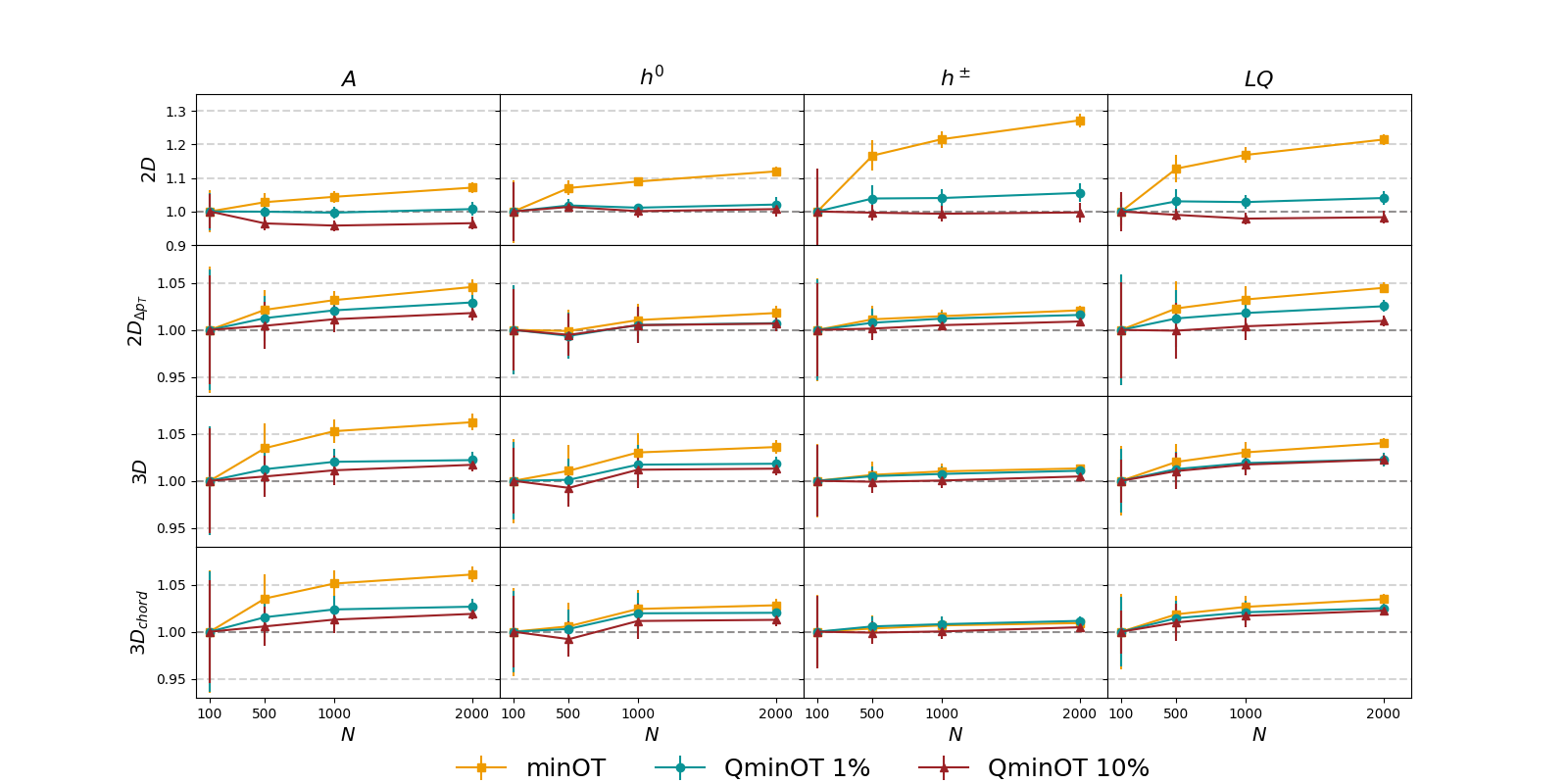}
    \caption{Scaling of relative AUC performance of closest OT distance methods as a function of increasing reference and test size $N$. We can see that all methods, with the exception of those relying on the 2D ground space distance, have $\lesssim 5\%$ deviation of the average AUC value from the value at $N=100$.
    Moreover, the deviation is often within the initial uncertainty range on the AUC measurement at $N=100$.}
\label{fig:OTonlyAsFunctionOfN}
\end{figure}

\begin{figure}
    \centering
    \includegraphics[width=\linewidth]{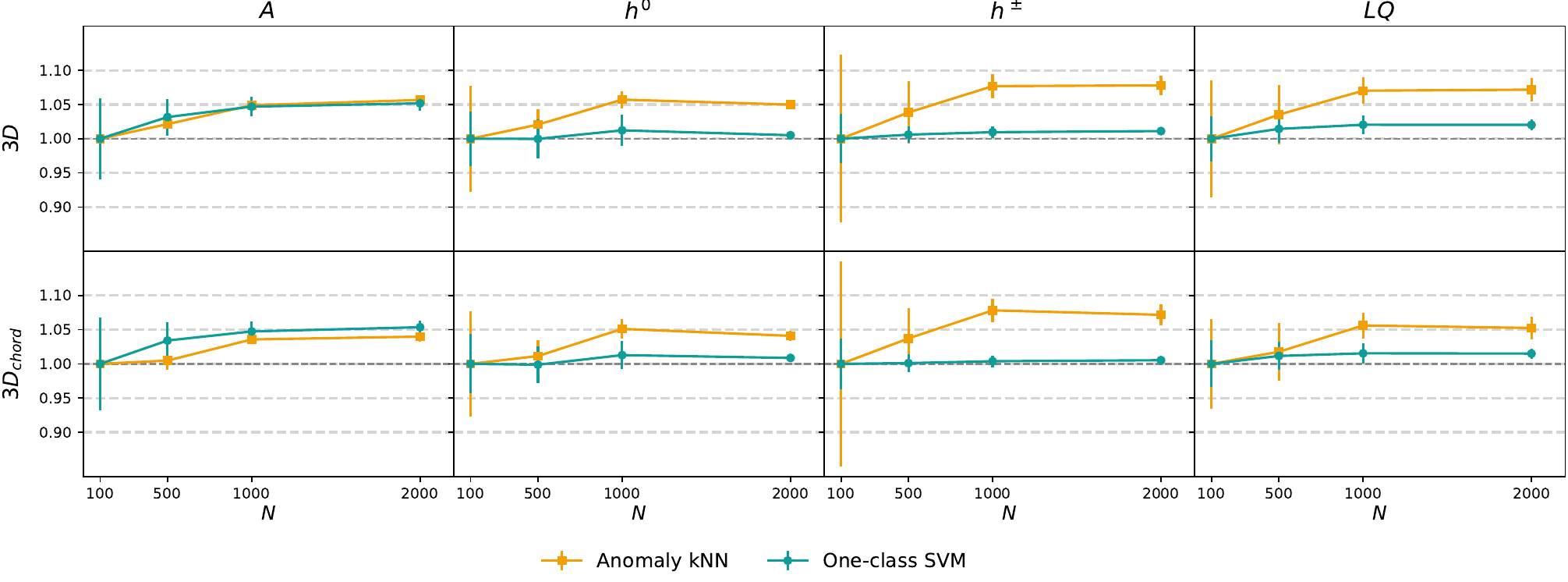}
    \caption{Scaling of relative AUC performance of OT+ML methods using 3D ground space distances as a function of increasing reference and test size $N$. We can see that both the OT+anomaly kNN and OT+One Class SVM are stable within the uncertainty band of the value at $N=100$.
    }
    \label{fig:OT+MLAsFunctionOfN}
\end{figure}

In this appendix, we consider how the performance of the various methods scale as the number of reference and test events ($N$) is increased. 
To probe this, we examine how the relative AUC changes as a function of $N$ relative to the method's performance at $N=100$: $\text{AUC}_{N}/\text{AUC}_{100}$.
Figure~\ref{fig:OTonlyAsFunctionOfN} examines this for OT-only methods for every choice of ground space distance considered. And Figure~\ref{fig:OT+MLAsFunctionOfN} examines this for the OT+ML methods using the 3D and 3D chord ground spaces. 

For the OT-only methods, we find that nearly all methods have less than a $5\%$ deviation of the average AUC value from the value at $N=100$. This deviation also falls within the initial uncertainty range on the AUC measurement at $N=100$. 
Thus, we can conclude that these OT-only methods are generally stable.
The exception to this is the minOT method with a 2D ground space distance. For certain signals, performance varies by $>20\%$. 
Overall, as expected, the $\text{QminOT}_p$ method is as stable or more stable than the minOT method for all ground space distances and signal cases.
For the OT+ML methods with 3D ground space distances, both are similarly stable within initial uncertainty ranges on the AUC measurement at $N=100$. 

\end{appendix}

\bibliography{references}

@misc{nguyen2023,
      title={Energy-Based Sliced Wasserstein Distance}, 
      author={Khai Nguyen and Nhat Ho},
      year={2023},
      eprint={2304.13586},
      archivePrefix={arXiv},
      primaryClass={stat.ML},
      url={https://arxiv.org/abs/2304.13586}, 
}

@misc{altschuler2018,
      title={Near-linear time approximation algorithms for optimal transport via Sinkhorn iteration}, 
      author={Jason Altschuler and Jonathan Weed and Philippe Rigollet},
      year={2018},
      eprint={1705.09634},
      archivePrefix={arXiv},
      primaryClass={cs.DS},
      url={https://arxiv.org/abs/1705.09634}, 
}

@misc{cuturi2013sinkhorndistanceslightspeedcomputation,
      title={Sinkhorn Distances: Lightspeed Computation of Optimal Transportation Distances}, 
      author={Marco Cuturi},
      year={2013},
      eprint={1306.0895},
      archivePrefix={arXiv},
      primaryClass={stat.ML},
      url={https://arxiv.org/abs/1306.0895}, 
}

@article{Bonneel2011,
author = {Bonneel, Nicolas and van de Panne, Michiel and Paris, Sylvain and Heidrich, Wolfgang},
title = {Displacement interpolation using Lagrangian mass transport},
year = {2011},
issue_date = {December 2011},
publisher = {Association for Computing Machinery},
address = {New York, NY, USA},
volume = {30},
number = {6},
issn = {0730-0301},
url = {https://doi.org/10.1145/2070781.2024192},
doi = {10.1145/2070781.2024192},
journal = {ACM Trans. Graph.},
month = dec,
pages = {1–12},
numpages = {12}
}

@article{Wang2013LinearOT,
  author    = {Wei Wang and Dejan Slep{\v{c}}ev and Saurav Basu and John A. Ozolek and Gustavo K. Rohde},
  title     = {A Linear Optimal Transportation Framework for Quantifying and Visualizing Variations in Sets of Images},
  journal   = {Int. J. Comput. Vis.},
  year      = {2012},
  volume    = {101},
  number    = {2},
  pages     = {254--269},
  doi       = {10.1007/s11263-012-0566-z},
  pmid      = {23729991},
  pmcid     = {PMC3667970},
  issn      = {1573-1405},
  publisher = {Springer},
}

@misc{kolouri2019generalizedslicedwassersteindistances,
      title={Generalized Sliced Wasserstein Distances}, 
      author={Soheil Kolouri and Kimia Nadjahi and Umut Simsekli and Roland Badeau and Gustavo K. Rohde},
      year={2019},
      eprint={1902.00434},
      archivePrefix={arXiv},
      primaryClass={cs.LG},
      url={https://arxiv.org/abs/1902.00434}, 
}

@article {sliced_wasserstein,
	title = {Sliced and Radon Wasserstein Barycenters of Measures},
	journal = {Journal of Mathematical Imaging and Vision},
	year = {2014},
	author = {Nicolas Bonneel and Julien Rabin and Gabriel Peyre and Hanspeter Pfister}
}

@article{Hyndman01111996,
author = {Rob J. Hyndman and Yanan Fan},
title = {Sample Quantiles in Statistical Packages},
journal = {The American Statistician},
volume = {50},
number = {4},
pages = {361--365},
year = {1996},
publisher = {Taylor \& Francis},
doi = {10.1080/00031305.1996.10473566},

URL = { 
https://www.tandfonline.com/doi/abs/10.1080/00031305.1996.10473566
},
eprint = {         
https://www.tandfonline.com/doi/pdf/10.1080/00031305.1996.10473566
}

}

@article{Farina:2018fyg,
    author = "Farina, Marco and Nakai, Yuichiro and Shih, David",
    title = "{Searching for New Physics with Deep Autoencoders}",
    eprint = "1808.08992",
    archivePrefix = "arXiv",
    primaryClass = "hep-ph",
    doi = "10.1103/PhysRevD.101.075021",
    journal = "Phys. Rev. D",
    volume = "101",
    number = "7",
    pages = "075021",
    year = "2020"
}

@article{Heimel:2018mkt,
    author = "Heimel, Theo and Kasieczka, Gregor and Plehn, Tilman and Thompson, Jennifer M.",
    title = "{QCD or What?}",
    eprint = "1808.08979",
    archivePrefix = "arXiv",
    primaryClass = "hep-ph",
    doi = "10.21468/SciPostPhys.6.3.030",
    journal = "SciPost Phys.",
    volume = "6",
    number = "3",
    pages = "030",
    year = "2019"
}

@article{Sjostrand:2014zea,
    author = {Sj\"ostrand, Torbj\"orn and Ask, Stefan and Christiansen, Jesper R. and Corke, Richard and Desai, Nishita and Ilten, Philip and Mrenna, Stephen and Prestel, Stefan and Rasmussen, Christine O. and Skands, Peter Z.},
    title = "{An introduction to PYTHIA 8.2}",
    eprint = "1410.3012",
    archivePrefix = "arXiv",
    primaryClass = "hep-ph",
    reportNumber = "LU-TP-14-36, MCNET-14-22, CERN-PH-TH-2014-190, FERMILAB-PUB-14-316-CD, DESY-14-178, SLAC-PUB-16122",
    doi = "10.1016/j.cpc.2015.01.024",
    journal = "Comput. Phys. Commun.",
    volume = "191",
    pages = "159--177",
    year = "2015"
}

@article{deFavereau:2013fsa,
    author = "de Favereau, J. and Delaere, C. and Demin, P. and Giammanco, A. and Lema\^\i{}tre, V. and Mertens, A. and Selvaggi, M.",
    collaboration = "DELPHES 3",
    title = "{DELPHES 3, A modular framework for fast simulation of a generic collider experiment}",
    eprint = "1307.6346",
    archivePrefix = "arXiv",
    primaryClass = "hep-ex",
    doi = "10.1007/JHEP02(2014)057",
    journal = "JHEP",
    volume = "02",
    pages = "057",
    year = "2014"
}

@article{Cacciari:2008gp,
    author = "Cacciari, Matteo and Salam, Gavin P. and Soyez, Gregory",
    title = "{The anti-$k_t$ jet clustering algorithm}",
    eprint = "0802.1189",
    archivePrefix = "arXiv",
    primaryClass = "hep-ph",
    reportNumber = "LPTHE-07-03",
    doi = "10.1088/1126-6708/2008/04/063",
    journal = "JHEP",
    volume = "04",
    pages = "063",
    year = "2008"
}

@article{Larkoski:2023qnv,
    author = "Larkoski, Andrew J. and Thaler, Jesse",
    title = "{A spectral metric for collider geometry}",
    eprint = "2305.03751",
    archivePrefix = "arXiv",
    primaryClass = "hep-ph",
    reportNumber = "MIT-CTP 5556",
    doi = "10.1007/JHEP08(2023)107",
    journal = "JHEP",
    volume = "08",
    pages = "107",
    year = "2023"
}

@article{Cai:2021hnn,
    author = "Cai, Tianji and Cheng, Junyi and Craig, Katy and Craig, Nathaniel",
    title = "{Which metric on the space of collider events?}",
    eprint = "2111.03670",
    archivePrefix = "arXiv",
    primaryClass = "hep-ph",
    doi = "10.1103/PhysRevD.105.076003",
    journal = "Phys. Rev. D",
    volume = "105",
    number = "7",
    pages = "076003",
    year = "2022"
}

@article{Cai:2020vzx,
    author = "Cai, Tianji and Cheng, Junyi and Craig, Nathaniel and Craig, Katy",
    title = "{Linearized optimal transport for collider events}",
    eprint = "2008.08604",
    archivePrefix = "arXiv",
    primaryClass = "hep-ph",
    doi = "10.1103/PhysRevD.102.116019",
    journal = "Phys. Rev. D",
    volume = "102",
    number = "11",
    pages = "116019",
    year = "2020"
}

@article{Komiske:2020qhg,
    author = "Komiske, Patrick T. and Metodiev, Eric M. and Thaler, Jesse",
    title = "{The Hidden Geometry of Particle Collisions}",
    eprint = "2004.04159",
    archivePrefix = "arXiv",
    primaryClass = "hep-ph",
    reportNumber = "MIT-CTP 5185",
    doi = "10.1007/JHEP07(2020)006",
    journal = "JHEP",
    volume = "07",
    pages = "006",
    year = "2020"
}

@article{Komiske:2019fks,
    author = "Komiske, Patrick T. and Metodiev, Eric M. and Thaler, Jesse",
    title = "{Metric Space of Collider Events}",
    eprint = "1902.02346",
    archivePrefix = "arXiv",
    primaryClass = "hep-ph",
    reportNumber = "MIT-CTP 5102",
    doi = "10.1103/PhysRevLett.123.041801",
    journal = "Phys. Rev. Lett.",
    volume = "123",
    number = "4",
    pages = "041801",
    year = "2019"
}

@article{Algren:2023spv,
    author = "Algren, Malte and Raine, John Andrew and Golling, Tobias",
    title = "{Decorrelation using Optimal Transport}",
    eprint = "2307.05187",
    archivePrefix = "arXiv",
    primaryClass = "hep-ph",
    month = "7",
    year = "2023"
}

@article{Cacciari:2011ma,
    author = "Cacciari, Matteo and Salam, Gavin P. and Soyez, Gregory",
    title = "{FastJet User Manual}",
    eprint = "1111.6097",
    archivePrefix = "arXiv",
    primaryClass = "hep-ph",
    reportNumber = "CERN-PH-TH-2011-297",
    doi = "10.1140/epjc/s10052-012-1896-2",
    journal = "Eur. Phys. J. C",
    volume = "72",
    pages = "1896",
    year = "2012"
}

@article{Gaertner:2023tzv,
    author = "Gaertner, Tejes and Reiten, Jared",
    title = "{Unsupervised learning in the metric space of jets}",
    eprint = "2312.06948",
    archivePrefix = "arXiv",
    primaryClass = "hep-ph",
    month = "12",
    year = "2023"
}

@article{Ba:2023hix,
    author = "Ba, Demba and Dogra, Akshunna S. and Gambhir, Rikab and Tasissa, Abiy and Thaler, Jesse",
    title = "{SHAPER: can you hear the shape of a jet?}",
    eprint = "2302.12266",
    archivePrefix = "arXiv",
    primaryClass = "hep-ph",
    reportNumber = "MIT-CTP 5535",
    doi = "10.1007/JHEP06(2023)195",
    journal = "JHEP",
    volume = "06",
    pages = "195",
    year = "2023"
}

@article{Davis:2023lxq,
    author = "Davis, Adam and Menzo, Tony and Youssef, Ahmed and Zupan, Jure",
    title = "{Earth mover\textquoteright{}s distance as a measure of CP violation}",
    eprint = "2301.13211",
    archivePrefix = "arXiv",
    primaryClass = "hep-ph",
    doi = "10.1007/JHEP06(2023)098",
    journal = "JHEP",
    volume = "06",
    pages = "098",
    year = "2023"
}

@article{scikit-learn,
 title={Scikit-learn: Machine Learning in {P}ython},
 author={Pedregosa, F. and Varoquaux, G. and Gramfort, A. and Michel, V.
         and Thirion, B. and Grisel, O. and Blondel, M. and Prettenhofer, P.
         and Weiss, R. and Dubourg, V. and Vanderplas, J. and Passos, A. and
         Cournapeau, D. and Brucher, M. and Perrot, M. and Duchesnay, E.},
 journal={Journal of Machine Learning Research},
 volume={12},
 pages={2825--2830},
 year={2011}
}

@article{flamary2021pot,
  author  = {R{\'e}mi Flamary and Nicolas Courty and Alexandre Gramfort and Mokhtar Z. Alaya and Aur{\'e}lie Boisbunon and Stanislas Chambon and Laetitia Chapel and Adrien Corenflos and Kilian Fatras and Nemo Fournier and L{\'e}o Gautheron and Nathalie T.H. Gayraud and Hicham Janati and Alain Rakotomamonjy and Ievgen Redko and Antoine Rolet and Antony Schutz and Vivien Seguy and Danica J. Sutherland and Romain Tavenard and Alexander Tong and Titouan Vayer},
  title   = {POT: Python Optimal Transport},
  journal = {Journal of Machine Learning Research},
  year    = {2021},
  volume  = {22},
  number  = {78},
  pages   = {1-8},
  url     = {http://jmlr.org/papers/v22/20-451.html}
}

@ARTICLE{1053964,
  author={Cover, T. and Hart, P.},
  journal={IEEE Transactions on Information Theory}, 
  title={Nearest neighbor pattern classification}, 
  year={1967},
  volume={13},
  number={1},
  pages={21-27},
  doi={10.1109/TIT.1967.1053964}}

@article{scholkopf1999support,
  title={Support vector method for novelty detection},
  author={Sch{\"o}lkopf, Bernhard and Williamson, Robert C and Smola, Alex and Shawe-Taylor, John and Platt, John},
  journal={Advances in neural information processing systems},
  volume={12},
  year={1999}
}

@article{memoli2011gromov,
  title={Gromov--Wasserstein distances and the metric approach to object matching},
  author={M{\'e}moli, Facundo},
  journal={Foundations of computational mathematics},
  volume={11},
  pages={417--487},
  year={2011},
  publisher={Springer}
}

@misc{li2023efficient,
      title={Efficient Approximation of Gromov-Wasserstein Distance Using Importance Sparsification}, 
      author={Mengyu Li and Jun Yu and Hongteng Xu and Cheng Meng},
      year={2023},
      eprint={2205.13573},
      archivePrefix={arXiv},
      primaryClass={cs.LG}
}

@article{Dillon:2023zac,
    author = "Dillon, Barry M. and Favaro, Luigi and Feiden, Friedrich and Modak, Tanmoy and Plehn, Tilman",
    title = "{Anomalies, Representations, and Self-Supervision}",
    eprint = "2301.04660",
    archivePrefix = "arXiv",
    primaryClass = "hep-ph",
    month = "1",
    year = "2023"
}

@article{Kasieczka:2022naq,
    author = "Kasieczka, Gregor and Mastandrea, Radha and Mikuni, Vinicius and Nachman, Benjamin and Pettee, Mariel and Shih, David",
    title = "{Anomaly detection under coordinate transformations}",
    eprint = "2209.06225",
    archivePrefix = "arXiv",
    primaryClass = "hep-ph",
    doi = "10.1103/PhysRevD.107.015009",
    journal = "Phys. Rev. D",
    volume = "107",
    number = "1",
    pages = "015009",
    year = "2023"
}

@article{Fraser:2021lxm,
    author = "Fraser, Katherine and Homiller, Samuel and Mishra, Rashmish K. and Ostdiek, Bryan and Schwartz, Matthew D.",
    title = "{Challenges for unsupervised anomaly detection in particle physics}",
    eprint = "2110.06948",
    archivePrefix = "arXiv",
    primaryClass = "hep-ph",
    doi = "10.1007/JHEP03(2022)066",
    journal = "JHEP",
    volume = "03",
    pages = "066",
    year = "2022"
}

@article{Govorkova:2021hqu,
    author = "Govorkova, Ekaterina and Puljak, Ema and Aarrestad, Thea and Pierini, Maurizio and Wo\'zniak, Kinga Anna and Ngadiuba, Jennifer",
    title = "{LHC physics dataset for unsupervised New Physics detection at 40 MHz}",
    eprint = "2107.02157",
    archivePrefix = "arXiv",
    primaryClass = "physics.data-an",
    reportNumber = "FERMILAB-PUB-21-338-CMS",
    doi = "10.1038/s41597-022-01187-8",
    journal = "Sci.  Data",
    volume = "9",
    pages = "118",
    year = "2022"
}

@article{Aarrestad:2021oeb,
    author = "Aarrestad, Thea and others",
    title = "{The Dark Machines Anomaly Score Challenge: Benchmark Data and Model Independent Event Classification for the Large Hadron Collider}",
    eprint = "2105.14027",
    archivePrefix = "arXiv",
    primaryClass = "hep-ph",
    reportNumber = "FERMILAB-PUB-21-285-CMS",
    doi = "10.21468/SciPostPhys.12.1.043",
    journal = "SciPost Phys.",
    volume = "12",
    number = "1",
    pages = "043",
    year = "2022"
}

@article{Kasieczka:2021xcg,
    author = "Kasieczka, Gregor and others",
    title = "{The LHC Olympics 2020 a community challenge for anomaly detection in high energy physics}",
    eprint = "2101.08320",
    archivePrefix = "arXiv",
    primaryClass = "hep-ph",
    doi = "10.1088/1361-6633/ac36b9",
    journal = "Rept. Prog. Phys.",
    volume = "84",
    number = "12",
    pages = "124201",
    year = "2021"
}

@article{Nachman:2020ccu,
    author = "Nachman, Benjamin",
    title = "{Anomaly Detection for Physics Analysis and Less than Supervised Learning}",
    eprint = "2010.14554",
    archivePrefix = "arXiv",
    primaryClass = "hep-ph",
    month = "10",
    year = "2020"
}

@misc{hepmllivingreview,  
    Author = "{HEP ML Community}",     
    title = "{A Living Review of Machine Learning for Particle Physics}",     
    url={https://iml-wg.github.io/HEPML-LivingReview/} }

@article{Belis:2023mqs,
    author = "Belis, Vasilis and Odagiu, Patrick and \r{A}rrestad, Thea Kl\ae{}boe",
    title = "{Machine Learning for Anomaly Detection in Particle Physics}",
    eprint = "2312.14190",
    archivePrefix = "arXiv",
    primaryClass = "physics.data-an",
    month = "12",
    year = "2023"
}

@article{Chang:2017kvc,
    author = "Chang, Spencer and Cohen, Timothy and Ostdiek, Bryan",
    title = "{What is the Machine Learning?}",
    eprint = "1709.10106",
    archivePrefix = "arXiv",
    primaryClass = "hep-ph",
    doi = "10.1103/PhysRevD.97.056009",
    journal = "Phys. Rev. D",
    volume = "97",
    number = "5",
    pages = "056009",
    year = "2018"
}

@article{Freytsis:2023cjr,
    author = "Freytsis, Marat and Perelstein, Maxim and San, Yik Chuen",
    title = "{Anomaly Detection in Presence of Irrelevant Features}",
    eprint = "2310.13057",
    archivePrefix = "arXiv",
    primaryClass = "hep-ph",
    month = "10",
    year = "2023"
}

@article{Fanelli:2023lmp,
    author = "Fanelli, Cristiano and Giroux, James",
    title = "{ELUQuant: Event-Level Uncertainty Quantification in Deep Inelastic Scattering}",
    eprint = "2310.02913",
    archivePrefix = "arXiv",
    primaryClass = "cs.LG",
    month = "10",
    year = "2023"
}

@article{Finke:2023ltw,
    author = {Finke, Thorben and Hein, Marie and Kasieczka, Gregor and Kr\"amer, Michael and M\"uck, Alexander and Prangchaikul, Parada and Quadfasel, Tobias and Shih, David and Sommerhalder, Manuel},
    title = "{Back To The Roots: Tree-Based Algorithms for Weakly Supervised Anomaly Detection}",
    eprint = "2309.13111",
    archivePrefix = "arXiv",
    primaryClass = "hep-ph",
    reportNumber = "TTK-23-26",
    month = "9",
    year = "2023"
}

@article{ATLAS:2023azi,
    author = "Aad, Georges and others",
    collaboration = "ATLAS",
    title = "{Anomaly detection search for new resonances decaying into a Higgs boson and a generic new particle $X$ in hadronic final states using $\sqrt{s} = 13$ TeV $pp$ collisions with the ATLAS detector}",
    eprint = "2306.03637",
    archivePrefix = "arXiv",
    primaryClass = "hep-ex",
    reportNumber = "CERN-EP-2023-045",
    month = "6",
    year = "2023"
}

@article{ATLAS:2020iwa,
    author = "Aad, Georges and others",
    collaboration = "ATLAS",
    title = "{Dijet resonance search with weak supervision using $\sqrt{s}=13$ TeV $pp$ collisions in the ATLAS detector}",
    eprint = "2005.02983",
    archivePrefix = "arXiv",
    primaryClass = "hep-ex",
    reportNumber = "CERN-EP-2020-062",
    doi = "10.1103/PhysRevLett.125.131801",
    journal = "Phys. Rev. Lett.",
    volume = "125",
    number = "13",
    pages = "131801",
    year = "2020"
}

@article{Govorkova:2021utb,
    author = "Govorkova, Ekaterina and others",
    title = "{Autoencoders on field-programmable gate arrays for real-time, unsupervised new physics detection at 40 MHz at the Large Hadron Collider}",
    eprint = "2108.03986",
    archivePrefix = "arXiv",
    primaryClass = "physics.ins-det",
    reportNumber = "FERMILAB-PUB-21-487-CMS, FERMILAB-PUB-21-487-CMS",
    doi = "10.1038/s42256-022-00441-3",
    journal = "Nature Mach. Intell.",
    volume = "4",
    pages = "154--161",
    year = "2022"
}

@inproceedings{Zipper:2023ybp,
    author = "Zipper, Noah",
    collaboration = "CMS",
    title = "{Testing a Neural Network for Anomaly Detection in the CMS Global Trigger Test Crate during Run 3}",
    booktitle = "{Topical Workshop on Electronics for Particle Physics}",
    eprint = "2312.10009",
    archivePrefix = "arXiv",
    primaryClass = "hep-ex",
    reportNumber = "CMS-CR-2023-255",
    month = "12",
    year = "2023"
}

@misc{asres2023spatiotemporal,
      title={Spatio-Temporal Anomaly Detection with Graph Networks for Data Quality Monitoring of the Hadron Calorimeter}, 
      author={Mulugeta Weldezgina Asres and Christian Walter Omlin and Long Wang and David Yu and Pavel Parygin and Jay Dittmann and Georgia Karapostoli and Markus Seidel and Rosamaria Venditti and Luka Lambrecht and Emanuele Usai and Muhammad Ahmad and Javier Fernandez Menendez and Kaori Maeshima and the CMS-HCAL Collaboration},
      year={2023},
      eprint={2311.04190},
      archivePrefix={arXiv},
      primaryClass={cs.LG}
}

@article{Grosso:2023owo,
    author = "Grosso, Gaia and Lai, Nicol\`o and Migliorini, Matteo and Pazzini, Jacopo and Triossi, Andrea and Zanetti, Marco and Zucchetta, Alberto",
    title = "{Triggerless data acquisition pipeline for Machine Learning based statistical anomaly detection}",
    eprint = "2311.02038",
    archivePrefix = "arXiv",
    primaryClass = "hep-ex",
    month = "11",
    year = "2023"
}

@article{CMSECAL:2023fvz,
    author = "Abadjiev, D. and others",
    collaboration = "CMS ECAL",
    title = "{Autoencoder-based Anomaly Detection System for Online Data Quality Monitoring of the CMS Electromagnetic Calorimeter}",
    eprint = "2309.10157",
    archivePrefix = "arXiv",
    primaryClass = "physics.ins-det",
    month = "9",
    year = "2023"
}

@inproceedings{Harilal:2023smf,
    author = "Harilal, Abhirami and Park, Kyungmin and Andrews, Michael and Paulini, Manfred",
    collaboration = "CMS",
    title = "{Autoencoder-based Online Data Quality Monitoring for the CMS Electromagnetic Calorimeter}",
    booktitle = "{21th International Workshop on Advanced Computing and Analysis Techniques in Physics Research}: {AI meets Reality}",
    eprint = "2308.16659",
    archivePrefix = "arXiv",
    primaryClass = "physics.ins-det",
    reportNumber = "CMS-CR-2023-020",
    month = "8",
    year = "2023"
}

@article{ATLAS:2023ixc,
    author = "Aad, Georges and others",
    collaboration = "ATLAS",
    title = "{Search for new phenomena in two-body invariant mass distributions using unsupervised machine learning for anomaly detection at $\sqrt{s} = 13$ TeV with the ATLAS detector}",
    eprint = "2307.01612",
    archivePrefix = "arXiv",
    primaryClass = "hep-ex",
    reportNumber = "CERN-EP-2023-112",
    month = "7",
    year = "2023"
}

@article{Chekanov:2023uot,
    author = "Chekanov, Sergei V. and Zhang, Rui",
    title = "{Boosting sensitivity to new physics with unsupervised anomaly detection in dijet resonance search}",
    eprint = "2308.02671",
    archivePrefix = "arXiv",
    primaryClass = "hep-ex",
    reportNumber = "ANL-HEP-183852",
    month = "8",
    year = "2023"
}

@article{Golling:2023yjq,
    author = "Golling, Tobias and Kasieczka, Gregor and Krause, Claudius and Mastandrea, Radha and Nachman, Benjamin and Raine, John Andrew and Sengupta, Debajyoti and Shih, David and Sommerhalder, Manuel",
    title = "{The Interplay of Machine Learning--based Resonant Anomaly Detection Methods}",
    eprint = "2307.11157",
    archivePrefix = "arXiv",
    primaryClass = "hep-ph",
    month = "7",
    year = "2023"
}

@article{Schuhmacher:2023pro,
    author = "Schuhmacher, Julian and Boggia, Laura and Belis, Vasilis and Puljak, Ema and Grossi, Michele and Pierini, Maurizio and Vallecorsa, Sofia and Tacchino, Francesco and Barkoutsos, Panagiotis and Tavernelli, Ivano",
    title = "{Unravelling physics beyond the standard model with classical and quantum anomaly detection}",
    eprint = "2301.10787",
    archivePrefix = "arXiv",
    primaryClass = "hep-ex",
    month = "1",
    year = "2023"
}

@article{Fanelli:2022xwl,
    author = "Fanelli, C. and Giroux, J. and Papandreou, Z.",
    title = "{\textquoteleft{}Flux+Mutability\textquoteright{}: a conditional generative approach to one-class classification and anomaly detection}",
    eprint = "2204.08609",
    archivePrefix = "arXiv",
    primaryClass = "cs.LG",
    doi = "10.1088/2632-2153/ac9bcb",
    journal = "Mach. Learn. Sci. Tech.",
    volume = "3",
    number = "4",
    pages = "045012",
    year = "2022"
}

@article{CrispimRomao:2020ejk,
    author = "Crispim Rom\~ao, M. and Castro, N. F. and Milhano, J. G. and Pedro, R. and Vale, T.",
    title = "{Use of a generalized energy Mover\textquoteright{}s distance in the search for rare phenomena at colliders}",
    eprint = "2004.09360",
    archivePrefix = "arXiv",
    primaryClass = "hep-ph",
    doi = "10.1140/epjc/s10052-021-08891-6",
    journal = "Eur. Phys. J. C",
    volume = "81",
    number = "2",
    pages = "192",
    year = "2021"
}

@article{Mikuni:2023tok,
    author = "Mikuni, Vinicius and Nachman, Benjamin",
    title = "{High-dimensional and Permutation Invariant Anomaly Detection}",
    eprint = "2306.03933",
    archivePrefix = "arXiv",
    primaryClass = "hep-ph",
    month = "6",
    year = "2023"
}

@article{Park:2022zov,
    author = "Park, Sang Eon and Harris, Philip and Ostdiek, Bryan",
    title = "{Neural embedding: learning the embedding of the manifold of physics data}",
    eprint = "2208.05484",
    archivePrefix = "arXiv",
    primaryClass = "hep-ph",
    doi = "10.1007/JHEP07(2023)108",
    journal = "JHEP",
    volume = "07",
    pages = "108",
    year = "2023"
}

@article{Dillon:2022mkq,
    author = {Dillon, Barry M. and Favaro, Luigi and Plehn, Tilman and Sorrenson, Peter and Kr\"amer, Michael},
    title = "{A Normalized Autoencoder for LHC Triggers}",
    eprint = "2206.14225",
    archivePrefix = "arXiv",
    primaryClass = "hep-ph",
    month = "6",
    year = "2022"
}

@article{Finke:2021sdf,
    author = {Finke, Thorben and Kr\"amer, Michael and Morandini, Alessandro and M\"uck, Alexander and Oleksiyuk, Ivan},
    title = "{Autoencoders for unsupervised anomaly detection in high energy physics}",
    eprint = "2104.09051",
    archivePrefix = "arXiv",
    primaryClass = "hep-ph",
    reportNumber = "TTK-21-12",
    doi = "10.1007/JHEP06(2021)161",
    journal = "JHEP",
    volume = "06",
    pages = "161",
    year = "2021"
}

@article{Dillon:2021nxw,
    author = "Dillon, Barry M. and Plehn, Tilman and Sauer, Christof and Sorrenson, Peter",
    title = "{Better Latent Spaces for Better Autoencoders}",
    eprint = "2104.08291",
    archivePrefix = "arXiv",
    primaryClass = "hep-ph",
    doi = "10.21468/SciPostPhys.11.3.061",
    journal = "SciPost Phys.",
    volume = "11",
    pages = "061",
    year = "2021"
}

@article{Mikuni:2021nwn,
    author = "Mikuni, Vinicius and Nachman, Benjamin and Shih, David",
    title = "{Online-compatible unsupervised nonresonant anomaly detection}",
    eprint = "2111.06417",
    archivePrefix = "arXiv",
    primaryClass = "cs.LG",
    doi = "10.1103/PhysRevD.105.055006",
    journal = "Phys. Rev. D",
    volume = "105",
    number = "5",
    pages = "055006",
    year = "2022"
}

@article{Batson:2021agz,
    author = "Batson, Joshua and Haaf, C. Grace and Kahn, Yonatan and Roberts, Daniel A.",
    title = "{Topological Obstructions to Autoencoding}",
    eprint = "2102.08380",
    archivePrefix = "arXiv",
    primaryClass = "hep-ph",
    reportNumber = "MIT-CTP/5264",
    doi = "10.1007/JHEP04(2021)280",
    journal = "JHEP",
    volume = "04",
    pages = "280",
    year = "2021"
}

@misc{gong2019memorizing,
      title={Memorizing Normality to Detect Anomaly: Memory-augmented Deep Autoencoder for Unsupervised Anomaly Detection}, 
      author={Dong Gong and Lingqiao Liu and Vuong Le and Budhaditya Saha and Moussa Reda Mansour and Svetha Venkatesh and Anton van den Hengel},
      year={2019},
      eprint={1904.02639},
      archivePrefix={arXiv},
      primaryClass={cs.CV}
}

@article{higgs_CMS,
   title={Observation of a new boson at a mass of 125 GeV with the CMS experiment at the LHC},
   volume={716},
   ISSN={0370-2693},
   url={http://dx.doi.org/10.1016/j.physletb.2012.08.021},
   DOI={10.1016/j.physletb.2012.08.021},
   number={1},
   journal={Physics Letters B},
   publisher={Elsevier BV},
   author={Chatrchyan, S. and Khachatryan, V. and Sirunyan, A.M. and Tumasyan, A. and Adam, W. and Aguilo, E. and Bergauer, T. and Dragicevic, M. and Erö, J. and Fabjan, C. and et al.},
   year={2012},
   month={Sep},
   pages={30–61}
}

@article{higgs_ATLAS,
   title={Observation of a new particle in the search for the Standard Model Higgs boson with the ATLAS detector at the LHC},
   volume={716},
   ISSN={0370-2693},
   url={http://dx.doi.org/10.1016/j.physletb.2012.08.020},
   DOI={10.1016/j.physletb.2012.08.020},
   number={1},
   journal={Physics Letters B},
   publisher={Elsevier BV},
   author={Aad, G. and Abajyan, T. and Abbott, B. and Abdallah, J. and Abdel Khalek, S. and Abdelalim, A.A. and Abdinov, O. and Aben, R. and Abi, B. and Abolins, M. and et al.},
   year={2012},
   month={Sep},
   pages={1–29}
}

@article{toappear,
	author = "Cai, Tianji and Craig, Katy and Craig, Nathaniel and Lin, Xinyuan",
	title = "to appear",
	year = "2024"
}

\nolinenumbers

\end{document}